\documentclass{aa}  
\usepackage[super]{nth}
\usepackage{graphicx}
\usepackage{txfonts}
\usepackage{lscape}
\usepackage{float}
\usepackage{natbib}
\usepackage{hyperref} 

\usepackage{xspace}
\usepackage{xcolor}
\usepackage{xfrac}
\usepackage{array,multirow}
\usepackage{placeins}

\makeatletter
\renewcommand*\aa@pageof{, page \thepage{} of \pageref*{LastPage}}
\makeatother

\bibpunct{(}{)}{;}{a}{}{,} 

\usepackage{tikz}
\definecolor{lime}{HTML}{A6CE39}
\DeclareRobustCommand{\orcidicon}{
	\begin{tikzpicture}
	\draw[lime, fill=lime] (0,0) 
	circle [radius=0.16] 
	node[white] {{\fontfamily{qag}\selectfont \tiny ID}};
	\draw[white, fill=white] (-0.0625,0.095) 
	circle [radius=0.007];
	\end{tikzpicture}
	\hspace{-2mm}}

\foreach \x in {A, ..., Z}{\expandafter\xdef\csname orcid\x
\endcsname{\noexpand\href{https://orcid.org/\csname orcidauthor\x\endcsname}{\noexpand\orcidicon}}}

\newcommand{\vsini}{\mbox{$v\sin i$}\xspace}
\newcommand{\vmac} {$v_{\rm mac}$\xspace}
\newcommand{\vmic}{$\xi$\xspace}

\newcommand{\Teff}{\mbox{$T_{\rm eff}$}\xspace}
\newcommand{\logg}{\mbox{$\log g$}\xspace}

\newcommand{\logQ}{\mbox{$\log Q$}\xspace}
\newcommand{\logLs}{$\log (\mathcal{L}/\mathcal{L}_{\odot})$\xspace}
\newcommand{\Ls}{$ \mathcal{L}$\xspace}
\newcommand{\He}{$Y_{\rm He}$\xspace}
\newcommand{\fwhb}{{\tt FW3414\,}(H$\beta$)\xspace}

\let\oldAA\AA
\renewcommand*{\AA}{\,\oldAA\xspace}
\newcommand{\kms}{\,\mbox{km\,s$^{-1}$}\xspace}
\newcommand{\MSol}{\,\mbox{M$_\odot$}\xspace}

\newcommand{\ls}{\mbox{$\lesssim$}\,}
\newcommand{\gs}{\mbox{$\gtrsim$}\,}

\begin{document}

\title{The IACOB project}
\subtitle{X.~Large-scale quantitative spectroscopic analysis of Galactic luminous blue stars}
\titlerunning{Large-scale quantitative spectroscopic analysis of Galactic luminous blue stars}
\author{de~Burgos, A.\inst{1,2\orcidA{}}, Simón-Díaz, S.\inst{1,2\orcidB{}}, Urbaneja, M.~A.\inst{3\orcidC{}}, Puls, J.\inst{4\orcidC{}}}
\authorrunning{A. de~Burgos et al.}
\institute{
Universidad de La Laguna, Dpto. Astrof\'isica, E-38206 La Laguna, Tenerife, Spain
\and
Instituto de Astrof\'isica de Canarias, Avenida V\'ia L\'actea, E-38205 La Laguna, Tenerife, Spain
\and
Universit\"at Innsbruck, Institut f\"ur Astro- und Teilchenphysik, Technikerstr. 25/8, A-6020 Innsbruck, Austria
\and
LMU M\"unchen, Universit\"atssternwarte, Scheinerstr. 1, 81679 M\"unchen, Germany
}
\date{Received 30 November 2023 / Accepted ---}
\abstract 
{Blue supergiants (BSGs) are key objects for understanding the evolution of massive stars, which play a crucial role in the evolution of galaxies. However, discrepancies between theoretical predictions and empirical observations have opened up important questions yet to be answered. Studying statistically significant and unbiased samples of these objects can help to improve the situation.
}
{To perform a homogeneous and comprehensive quantitative spectroscopic analysis of a large sample of Galactic luminous blue stars (a majority of which are BSGs) from the IACOB spectroscopic database, providing crucial parameters to refine and improve theoretical evolutionary models.
}
{We derive the projected rotational velocity (\vsini) and macroturbulent broadening (\vmac) using {\tt IACOB-BROAD}, which combines Fourier transform and line-profile fitting techniques. We compare high-quality optical spectra with state-of-the-art simulations of massive star atmospheres computed with the {\tt FASTWIND} code. This comparison allows us to derive effective temperatures (\Teff), surface gravities (\logg), microturbulences (\vmic), surface abundances of silicon and helium, and to assess the relevance of stellar winds through a wind-strength parameter (\logQ).
}
{We provide estimates and associated uncertainties of the above-mentioned quantities for the largest sample of Galactic luminous O9 to B5 stars spectroscopically analyzed to date, comprising 527 targets.
We find a clear drop in the relative number of stars at \Teff$\approx$\,21\,kK, coinciding with a scarcity of fast rotating stars below that temperature. We speculate that this feature (roughly corresponding to B2 spectral type) might be roughly delineating the location of the empirical Terminal-Age-Main-Sequence in the mass range between 15 and 85\MSol.
By investigating the main characteristics of the \vsini distribution of O stars and BSGs as a function of \Teff, we propose that an efficient mechanism transporting angular momentum from the stellar core to the surface might be operating along the main sequence in the high-mass domain.
We find correlations between \vmic, \vmac, and the spectroscopic luminosity \Ls (defined as $T_{\rm eff}^4$\,/$g$). We also find that no more than 20\% of the stars in our sample have atmospheres clearly enriched in helium, and suggest that the origin of this specific sub-sample might be in binary evolution. 
We do not find clear empirical evidence of an increase in the wind-strength over the wind bi-stability region towards lower \Teff.
}
%
{}

\maketitle


\section{Introduction}
\label{section:1_tmp}

Massive stars (M$_{ini}$\,$\gs$8\MSol) play a pivotal role in galactic systems, exerting a profound impact on their chemo-dynamical evolution. On the one hand, massive stars make a substantial contribution to the chemical enrichment of galaxies, primarily through supernova explosions, but also through the release of enriched material via stellar winds \citep[e.g.][]{maeder81, woosley95, kaufer97, nomoto13}. On the other hand, their dynamic influence extends to the surrounding interstellar medium, driven by intense stellar winds and the copious emission of UV radiation. These factors can profoundly shape the interstellar environment \citep[e.g.][]{krause13, watkins19, kim19, geen21}, either triggering or inhibiting new episodes of star formation.

These stars are intricately connected to some of the most energetic and dynamic phenomena in the Universe, such as core-collapse supernovae and gamma-ray bursts \citep{woosley06, smartt09}. Additionally, their role has recently attracted attention in the realm of gravitational-wave astrophysics, as they serve as progenitors of black hole and neutron star mergers \citep{abbott16, belczynski16, marchant16}.

Furthermore, massive stars are valuable tools for extragalactic research, serving as increasingly reliable distance indicators \citep{urbaneja17, taormina20} and providing unique insights into the present-day abundances of their host galaxies \citep{bresolin07, kudritzki12, bresolin16}, even at distances spanning several megaparsecs \citep{kudritzki03a, urbaneja03, urbaneja05a, kudritzki08, bresolin22}.

Blue supergiants (BSGs), a subset of massive stars, hold a crucial position in unraveling and understanding the intricate puzzle of the evolution of stars born with masses exceeding $\approx$15\MSol. (For a comprehensive overview of historical research and methodologies related to the study of BSGs, we refer to the introduction of a recent study by \citeauthor{webmayer22} \citeyear{webmayer22}.) Traditionally, BSGs were considered helium-burning stars that had completed their mainsequence (MS) evolution as single stars \citep[e.g.,][]{hayashi62}. However, decades of observations have revealed persistent discrepancies with theoretical models, indicating that the evolutionary status of BSGs is much more intricate \citep[see, e.g.,][]{fitzpatrick90, castro14, wang20}. This complexity likely arises from a range of diverse evolutionary pathways that can ultimately populate the region on the Hertzsprung-Russell diagram where BSGs are located \citep[see][]{vink10, maeder12, langer12}. In this context, the compilation and analysis of spectroscopic data of BSGs with a considerable increase in quality and sample size compared to previous works \citep[e.g.,][]{dufton72, lennon92, crowther06, lefever07, searle08, markova08, castro14, haucke18, webmayer22} is becoming an urgent need to decipher a more complex scenario than the one initially established.

Focused on this and related aspects, the IACOB project\footnote{\href{https://research.iac.es/proyecto/iacob/pages/en/introduction.php}{Link to the website of the IACOB project.}} started in 2008 with the overarching objective of providing high-quality empirical information on a statistically significant unbiased sample of Galactic massive stars, aiming to establish new anchor points for testing and improving current theories of stellar atmospheres, winds, interiors, and evolution of massive stars. Previous efforts of the IACOB team have mostly concentrated on the study of line-broadening sources affecting the spectra of O- and B-type stars \citep[][]{simon-diaz14a, simon-diaz14b, simon-diaz17, godart17} and the empirical characterization of Galactic targets covering the O star domain \citep[][]{holgado18, holgado20, holgado22, britavskiy23}.

Within this framework, the study presented in this paper, which can be considered a continuation of \citet{deburgos23}, aims at performing a homogeneous estimation of the relevant spectroscopic parameters of the most extensive sample of Galactic luminous blue stars compiled to date, with a specific focus on BSGs with O9 to B5 spectral types. In forthcoming papers, we will complement the results presented here with additional information on the luminosities, masses, radii, and surface abundances of key elements such as silicon, carbon, nitrogen, and oxygen, to cover other important quantities defining the properties of the sample. Our ultimate objective is to establish a new, highly improved, empirical standard for the study of these stellar objects.

The paper is organized as follows. Section~\ref{section:2_tmp} presents the spectroscopic dataset and the sample of stars under study. Section~\ref{section:3_tmp} describes the methodology used to obtain estimates for the line-broadening and other relevant spectroscopic parameters. Section~\ref{section:4_tmp} summarizes the results of the analysis and compares them with previous studies. In Sect.~\ref{section:5_tmp} we discuss the results of the analysis for the different parameters in our analysis, and in Sect.~\ref{section:6_tmp} we present the summary and conclusions of the work.


\section{Observational dataset and sample}
\label{section:2_tmp}

\begin{figure*}[t!]
\centering
\includegraphics[width=1\textwidth]{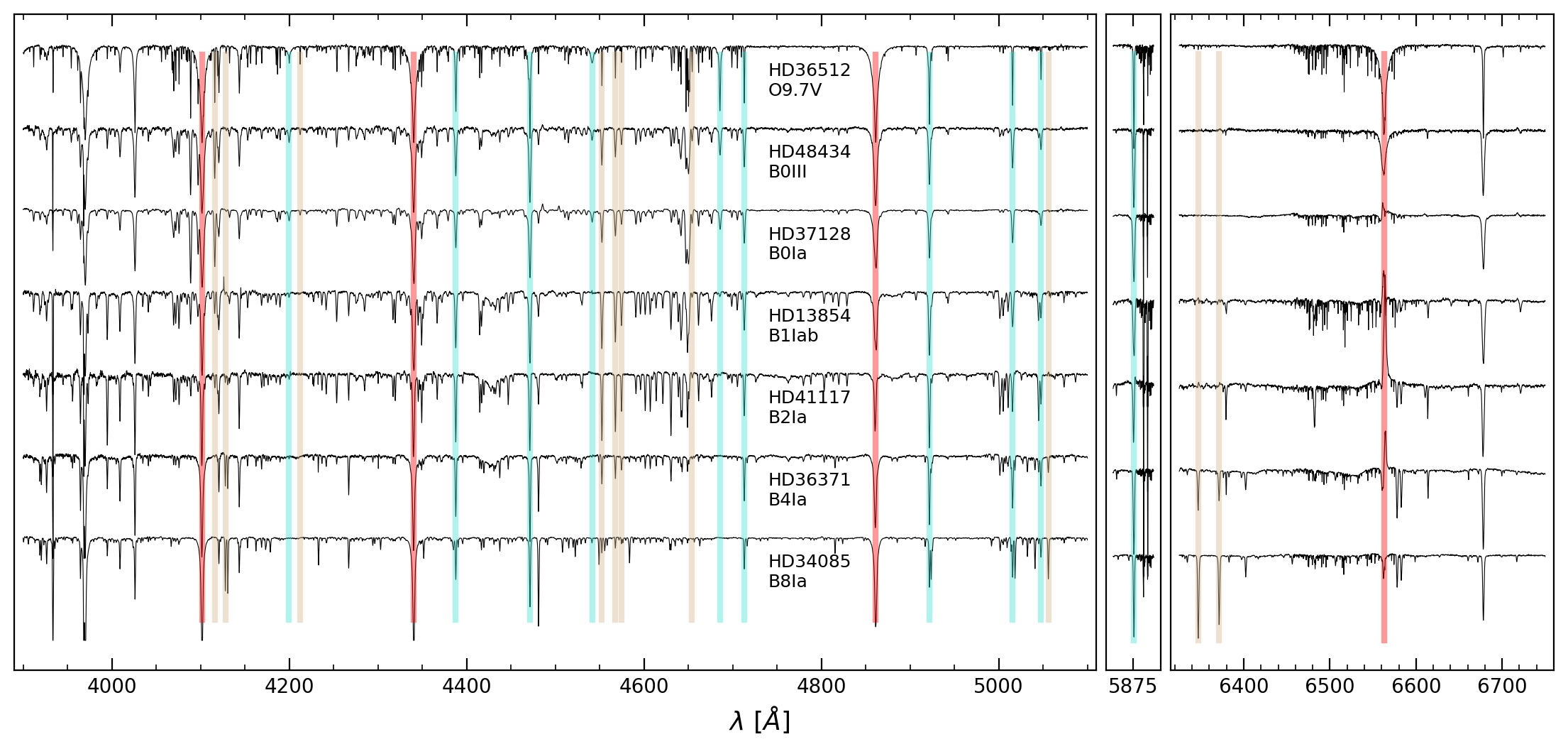}
\caption{Some illustrative examples of spectra used in this work, ordered by spectral type. Three different spectral windows depict the wavelength ranges in which the main diagnostic lines used to obtain estimates of the spectroscopic parameters are located. Vertical colored red, cyan, and brown bars indicate the corresponding H~{\sc i}, He~{\sc i-ii}, and Si~{\sc ii-iii-iv} lines, respectively (see Sect.~\ref{subsubsection:323_tmp} for further details).}
\label{fig:spectra}
\end{figure*}

This work makes use of the stellar sample described in \citet{deburgos23} and the associated spectroscopic data, which are collected from the IACOB spectroscopic database \citep[for the latest review see][]{simon-diaz20} and the ESO public archive. All considered spectra were obtained with FIES$@$NOT2.5m, HERMES$@$Mercator1.2m, and FEROS$@$MPG/ESO2.2m high-resolution echelle spectrographs that provide resolving powers between $R$\,25\,000 and $R$\,85\,000. The median signal-to-noise (S/N) ratio of the compiled dataset is $\approx$130 at 4500\,\AA. All spectra have a common wavelength coverage between 3800 and 7000\AA, reaching 9200\AA in some cases. Figure~\ref{fig:spectra} shows some examples of the quality of the spectroscopic observations that were analyzed in this work. 

Our original sample comprises 666 O9\,--\,B9 type stars selected from \citet{deburgos23}. In that work, we used the effect of gravity on the shape of H$\beta$ line as a proxy for the spectroscopic luminosity, \logLs (see Sect.~\ref{subsection:44_tmp}). Specifically, we used the quantity \fwhb, defined as the difference between the width of the H$\beta$ measured at three-quarters and one-quarter of its line depth, to select all O and B-type stars with initial masses above $\approx$20\MSol. This quantity represents an improvement over the traditional full width at half maximum (FWHM) by breaking the degeneracy caused by the projected rotational velocity (\vsini), while minimizing other effects such as surface temperature and spectral resolution \citep[see further notes in][]{deburgos23}. 

Concerning luminosity classes\footnote{See Table~\ref{tab:qsa_results} for information about the spectral classifications adopted in this work.} (LCs), our initial sample comprises 339 supergiants (LC\,I), 113 bright-giants (LC\,II), 111 giants (LC\,III), 55 subgiants (LC\,IV), and 48 dwarfs (LC\,V).

We note that the sample of 666 stars already excludes 12 B-type hypergiants, 27 classical Be-type stars \citep[see][and references therein]{negueruela04}, as well as 56 double-line or higher-order spectroscopic binaries (SB2+), identified in \citet{deburgos23}. This was done due to the impossibility of analyzing these objects with standard 1-D atmospheric models.


\section{Analysis methodology}
\label{section:3_tmp}

In this section, we describe the methodology used to derive the line-broadening and spectroscopic parameters of the stars in the sample. The analyses were carried out using the best available spectrum for each star (as quoted in Table~\ref{tab:qsa_results}), not only based on the S/N ratio but also regarding any potential issue affecting the spectral windows where the main diagnostic lines are located.


\subsection{Rotational and macroturbulent velocities}
\label{subsection:31_tmp}

Following \citet{simon-diaz14a}, we used {\tt IACOB-BROAD} to perform the line-broadening analysis. We obtained estimates of \vsini using the Fourier transform and the goodness-of-fit techniques. The latter also provided us with estimates of the macroturbulent velocity (\vmac, hereafter also referred to as {\em macroturbulence}). We checked the agreement of both techniques for deriving \vsini, and decided to keep the values from the goodness-of-fit as our final estimates. The few cases with larger differences ($\approx$20\kms) were attributed to low S/N spectra. Following \citet{deburgos23}, we used Si~{\sc iii} $\lambda$4567.85\,{\AA} and Si~{\sc ii} $\lambda$6371.37\,{\AA} for this analysis.


\subsection{Quantitative Spectroscopy}
\label{subsection:32_tmp}


\subsubsection{Model Atmosphere/Line formation code and main assumptions}
\label{subsubsection:321_tmp}

The NLTE model atmosphere/line synthesis code {\tt FASTWIND} \citep[Fast Analysis of STellar atmospheres with WINDs, v10.4.7,][]{santolaya-rey97, puls05, rivero-gonzalez11, puls20} was used to create a set of models for the analysis. A complete description of the current status of the code, as well as comparisons with alternative codes, have been presented by \citet{carneiro16}. {\tt FASTWIND} solves the radiative transfer problem in the comoving frame\footnote{For all lines from the so-called explicit elements that are used for the analysis (here hydrogen, helium, and silicon), as well as for the strongest lines from other elements (in between C and Zn); most other lines are treated within the Sobolev approximation.} of the expanding atmospheres of early-type stars in a spherically symmetric geometry, under the constraints of energy conservation and statistical equilibrium, and accounting for line-blocking/blanketing effects. Homogeneous chemical composition and steady state are also assumed. The density stratification is derived from the hydrostatic balance in the lower atmosphere, and from the mass-loss rate and the wind-velocity field (a standard $\beta$-law) via the equation of continuity in the wind. A smooth transition between the wind regime and the pseudo-static photosphere is enforced.

Each {\tt FASTWIND} simulation is defined by a set of parameters: the effective temperature (\Teff), surface gravity (\logg), and stellar radius ($R$), which are defined at $\tau_{\rm Ross} = 2/3$, the microturbulent velocity (\vmic), the exponent of the wind-velocity law ($\beta$), the mass-loss rate ($\dot{M}$), the wind terminal velocity (v$_\infty$), and a set of elemental chemical abundances.

Regarding any specific information concerning the detailed model atoms used in our calculations, we refer the reader to \citet{urbaneja05b}.


\subsubsection{Grid of model atmospheres}
\label{subsubsection:322_tmp}

\begin{table}[!t]
 \centering
 \caption{Parameter space covered by the model atmosphere calculations used in the present work.}
 \label{tab:param}
 \begin{tabular}{lcc}
 \hline
 \hline
    \noalign{\smallskip}
    Parameter & Abbreviation & Covered range \\ 
    \noalign{\smallskip}
    \hline
    \noalign{\smallskip}
    Effective temperature   & \Teff & 35\,--\,14\,kK \\   
    \noalign{\smallskip}
    Surface gravity         & \logg & 1.7\,--\,4.14\,dex \\
    \noalign{\smallskip}
    Microturbulence         & \vmic & 0\,--\,30\kms \\
    \noalign{\smallskip}
    Helium abundance$^{1}$  & \He   & 0.10\,--\,0.30 \\
    \noalign{\smallskip}
    Silicon abundance$^{2}$ & $\epsilon_{\rm Si}$ & 7.00\,--\,8.00 \\
    \noalign{\smallskip}
    Wind-strength$^{3}$     & $-$\logQ & 14.0\,--\,12.5 \\
    \noalign{\smallskip}
    Mass-loss rate$^{4}$    & $\dot{M}$ & (6.10\,--\,0.02)$\times10^{-6}$ \\
    \noalign{\smallskip}
    Wind terminal velocity  & v$_\infty$ & 2700\,--\,570\kms \\
    \noalign{\smallskip}
    Exponent of the wind    & $\beta$ & 0.8\,--\,3.0 \\
     velocity law           & & \\
    \noalign{\smallskip}\hline
 \end{tabular}
 \tablefoot{$^{1}\,Y_{\rm He} = \frac{\rm N(He)}{\rm N(H)}$. $^{2}\,\epsilon_{\rm Si} = \texttt{12+}\log\left(\frac{\rm N(Si)}{\rm N(H)}\right)$. $^{3}$\,\logQ values calculated in units of $M_\odot/a$, \kms and $R_\odot$. $^{4}$\, In units of $M_\odot/a$. }
\end{table}

As described in the previous section, each {\tt FASTWIND} model requires a set of seven parameters (plus elemental abundances). However, the optical spectrum of typical B-type supergiants (such as those analyzed in this work) does not contain relevant information that would allow us to constrain all these parameters in parallel. For example, the main signature of the stellar wind is imprinted into the H$\alpha$ profile, which for the most part is sensitive to the shape of the velocity field (i.e., $\beta$) and the wind-strength parameter (optical depth invariant) $Q$, a combination of mass-loss rate, wind terminal velocity and stellar radius\footnote{$Q$ is defined $\dot{M} / (R_{\star}v_{\infty})^{1.5}$ \citep[see][]{puls96, puls05}}, and not to the individual values of these three physical parameters. Nothing can be said about possible inhomogeneities likely to be present in the outflow, since only H$\alpha$, a recombination line, is available. Based on these considerations, we decided to consider only homogeneous winds (i.e., without clumping) since at the very least they will provide an upper limit for the wind strength via the wind-strength parameter $Q$. 

In consequence, each model in our grid is defined by a set of seven parameters: \Teff, \logg, \vmic, $\beta$, $Q$, helium and silicon abundances. The range covered by each parameter is indicated in Table~\ref{tab:param}). In addition, 
$(a)$ following \citet{urbaneja11} we used a fixed 10\kms microturbulence for the calculation of the atmospheric structure and occupation numbers but allowed for different (depth-independent) microturbulences for the calculation of the line profiles (formal solutions); 
$(b)$ we selected the lower limit of the effective temperature based on the fact that our models have not been thoroughly tested below \Teff$\approx$\,14\,kK;
$(c)$ the lower boundary for the helium abundance was selected to be the solar value \citep{magg22}, \He=\,N(He)/N(H)\,=\,0.10, as typically adopted in studies of Galactic massive stars; and 
$(d)$ all other elements beyond helium and silicon are adopted to follow the solar metallicities as in \citet{asplund09}.

To avoid the computation of an extremely large number of models that would be required in a classic regularly-spaced grid (reaching $\approx$1.5$\times$10$^{6}$ if we consider typical step-sizes in the sampling of the various considered parameters\footnote{In particular: 1000\,K in \Teff, 0.1\,dex in \logg, 5\kms in \vmic, 0.04 in \He, 0.15 in $\epsilon_{\rm Si}$, 0.3 in \logQ, and 0.7 in $\beta$.}), we opted for sampling the multi-D parameter space with a distribution of points following a Latin Hypercube Sampling algorithm \citep[LHS;][see also Appendix~\ref{apen.input_maui}]{mckay79, wei-liem96}. The resulting analysis grid comprises 358 {\tt FASTWIND} models ($\approx$55 per dimension in the parameter space). Using supervised learning techniques, these models are employed to train a statistical emulator \citep{mackay03}. This emulator is capable of reproducing {\tt FASTWIND} simulations to a specific degree of fidelity in a fraction of the time required to run any actual simulation (Urbaneja, M.A., in prep.). Later on, during the inference phase (see ~\ref{subsubsection:324_tmp}), this emulator is utilized in combination with a Metropolis-Hasting algorithm \citep{metropolis53} to sample from the underlying probability distribution.


\subsubsection{Diagnostic lines}
\label{subsubsection:323_tmp}

Table~\ref{tab:diag_lines} compiles the list of diagnostic lines chosen for this work. These were selected to be present in the 4000\,--\,7000\AA wavelength range, common to all the available spectra. Their location is shown in Fig.~\ref{fig:spectra} with different colors depending on the atomic element. 

These lines represent a minimum set required to obtain information on the fundamental atmospheric parameters characterizing the atmospheres of B-type supergiant stars. In particular, \Teff is inferred from the ionization balances He~{\sc i/ii} and Si~{\sc ii/iii/iv} (see for example \citealt{mcerlean99, urbaneja05b}). The hydrogen Balmer lines provide a strong constraint on \logg, due to their sensitivity to broadening via the Stark effect. When the stellar wind becomes strong enough, the shape and strength of the H$\alpha$ profile can provide constraints simultaneously on the wind acceleration $\beta$ and the wind-strength parameter $Q$.

The microturbulent velocity is estimated from the differential response of the three components of the strong Si~{\sc iii} $\lambda\lambda$ 4553-68-75 {\AA} triplet. We also note that some He~{\sc i} lines could show some sensitivity to this parameter \citep{mcerlean99}. However, the differential effect in the Si~{\sc iii} lines is the dominant source of information. Finally, the surface abundances are determined from the strength of the corresponding spectral lines of each species.

Strictly speaking, however, all the spectral features can (and will) react to more than one of the fundamental stellar parameters. For example, the hydrogen Balmer lines are also sensitive to \Teff and the helium abundance, albeit to a lower degree than to \logg. Similarly, the helium lines do not only depend on \Teff and helium abundance but also on \logg and microturbulence. Therefore, the analysis methodology involves a multi-dimensional optimization problem, in which the best solution is found in an iterative process, assuring at the same time a proper exploration of the full parameter space (see below).

Besides the physical arguments, when selecting lines, we avoided choosing those that are affected by known issues, such as, for example, showing blends with atomic species not currently included in our detailed model atoms, as well as lines that are severely affected by the presence of telluric lines. For example, the H$\epsilon$ line was excluded due to contamination with the strong interstellar calcium line.

\begin{table}[!t]
 \centering
 \caption{List of diagnostic lines used for the determination of the spectroscopic parameters.}
 \label{tab:diag_lines}
 \begin{tabular}{cc|cc|cc}
 \hline
 \hline
    \noalign{\smallskip}
    Line & $\lambda$ [\AA] & Line & $\lambda$ [\AA] & Line & $\lambda$ [\AA] \\ 
    \noalign{\smallskip}
 \hline
    \noalign{\smallskip}
    H$\delta$ & 4101.74 & He~{\sc i} & 4387.93 & He~{\sc ii} & 4199.83 \\ 
    H$\gamma$ & 4340.46 & He~{\sc i} & 4471.47 & He~{\sc ii} & 4541.59 \\ 
    H$\beta$  & 4861.33 & He~{\sc i} & 4713.14 & He~{\sc ii} & 5411.52  \\ 
    H$\alpha$ & 6562.80 & He~{\sc i} & 4921.93 &             &          \\ 
              &         & He~{\sc i} & 5015.68 &             &          \\ 
              &         & He~{\sc i} & 5047.74 &             &          \\ 
              &         & He~{\sc i} & 5875.62 &             &          \\ 
    \noalign{\smallskip}
 \hline
    \noalign{\smallskip}
    Si~{\sc ii} & 4128.05 & Si~{\sc iii} & 4552.62 & Si~{\sc iv} & 4116.10 \\
    Si~{\sc ii} & 4130.89 & Si~{\sc iii} & 4567.84 & Si~{\sc iv} & 4212.41 \\
    Si~{\sc ii} & 5056.32 & Si~{\sc iii} & 4574.76 & Si~{\sc iv} & 4654.31 \\
    Si~{\sc ii} & 6371.37 &  &  &  &  \\
 \hline
    \noalign{\smallskip}
 \end{tabular}
 \tablefoot{
 For reference, we found that the He~{\sc ii} and Si~{\sc iv} lines disappear below \Teff$\approx$\,25\,kK and \Teff$\approx$\,20\,kK, respectively. The He~{\sc i} and Si~{\sc iii} lines are present in the full \Teff range, and the Si~{\sc ii} lines appear below \Teff$\approx$\,21\,kK.
 H$\delta$ was removed for fast-rotating stars (see Sect.~\ref{subsubsection:323_tmp}).
 Despite the relatively strong Si~{\sc iv} 4088.96 line is also present, we decided to exclude it due to a blend with O~{\sc ii} $\lambda$4089.29\,{\AA}.}
\end{table}

Spectral features that show systematic differences between models and observations, suspected of suffering from modelling issues, were also excluded from the beginning. This is the case for the prominent He~{\sc i} $\lambda$6678\,{\AA} line, for which our models always predicted narrower lines than observed, which suggests that our current broadening data for this particular line are not fully adequate for B-type supergiant stars (see also Sect.~\ref{subsubsection:325_tmp} for other less important issues).

\begin{figure*}[t!]
\centering
\includegraphics[width=0.9\textwidth]{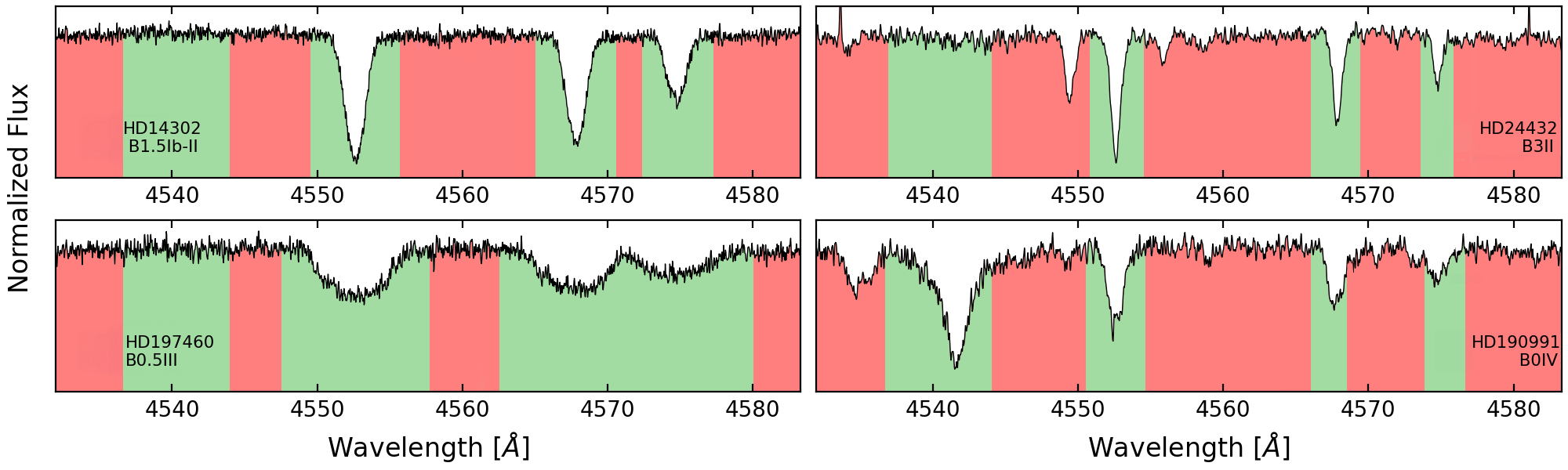}
\caption{Examples of different masks used to select the diagnostic lines. Each of the four panels shows the same wavelength range including the He~{\sc ii} $\lambda$4542\,{\AA} line and the Si~{\sc iii} $\lambda\lambda$4553,4568,4575\,{\AA} triplet. The two panels on the left compare two stars of similar temperature but different \vsini: HD\,14\,302 with \vsini$\approx$65\kms (top-left), and HD\,197\,460 with \vsini$\approx$200\kms (bottom-left). 
The two panels on the right compare two stars of similar \vsini but very different \Teff: HD\,24\,432 with \Teff=\,15\,kK (top-right) and HD\,190\,991 with \Teff=\,32\,kK (bottom-right). The wavelength ranges selected for each analysis are shaded in green. The masked (not used) regions are shaded in red.}
\label{fig:vsini_masks}
\end{figure*}

For each diagnostic line indicated in Table~\ref{tab:diag_lines}, the length of the corresponding spectral window used for the analysis was adjusted individually, taking into account the different intrinsic and rotational broadening. We also note that all the listed lines were always accounted for during the analysis process, even if a line was weak or not present. This is because such situations still provide important information (e.g., the absence of the He~{\sc ii} $\lambda$4542\,{\AA} -- see Fig.~\ref{fig:vsini_masks} -- indicates that the effective temperature of the star cannot exceed a certain value).

In addition, any contamination due to blends with lines from other species was masked out. We illustrate this in Fig.~\ref{fig:vsini_masks}, where we show the same spectral window for four different stars with their associated masks. It can be seen that the selected regions (in green) are different in all the cases, being more restrictive for the bottom-right panel, where multiple blends are present.


\subsubsection{Parameter inference}
\label{subsubsection:324_tmp}

The problem of determining the set of parameters $\overline{\pi}$ defining the model $M_\lambda$ that {\it best} reproduces an observation $O_\lambda$ can be mathematical described as finding the underlying probability distribution
\begin{align*}
p\left(\overline{\pi}, M_\lambda \mid O_\lambda\right) \propto p\left(\overline{\pi}, M_\lambda\right)\,p\left(O_\lambda \mid  \overline{\pi}, M_\lambda \right)
\end{align*}
where $p\left(\overline{\pi}, M_\lambda\right)$ represents the prior knowledge that we have on the models, and $p\left(O_\lambda \mid \overline{\pi}, M_\lambda \right)$ is the likelihood of an observation $O_\lambda$ given the model $M_\lambda$.

A well-proven method to sample from this unknown posterior distribution, to recover the ``best" parameter set and their corresponding uncertainties, is the use of a Markov chain Monte Carlo (MCMC) algorithm \citep{metropolis53, siddhartha01}. Key to the inference of the parameters is the definition of what is considered {\it best}, i.e. what criteria are used to evaluate the likelihood of a model, given an observed spectrum, as well as the prior knowledge on the parameter space. For the likelihood function (that is, the probability that a specific model $M_\lambda$ defined by a set of parameters $\overline{\pi}$ fits a given observed spectrum $O$) we adopt \citep{mackay03}
\begin{align*}
p\left(O_\lambda \mid \overline{\pi}, M_\lambda\right) \propto \exp{\left(-\chi^2 / 2\right)}
\end{align*}
where for each diagnostic window defined to contain the lines in Table~\ref{tab:diag_lines}, the merit function $\chi$ is defined as the sum of the quadratic residuals, weighted by the uncertainties, and normalized to the effective number of wavelength points $n_p$ contributing to the sum. Hence for each window, this corresponds to
\begin{align*}
\chi_{line}^2=\frac{1}{n_p} \sum_{j=1}^{n_p}\left(\frac{O_j-M_j}{\sigma_j}\right)^2
\end{align*}

Since in the construction of the Markov chain we are using emulated {\tt FASTWIND} spectra and not direct simulations (see Sect.~\ref{subsubsection:322_tmp}), we convinced ourselves that the level of accuracy obtained by the statistical emulator is good enough as not to affect the outcome of the analysis, i.e. that the possible uncertainties $\sigma_e$ introduced in the synthetic line profiles due to the statistical nature of the emulation are always significantly below the photon-noise level $\sigma_p$. Therefore $\sigma_\lambda^2 = \sigma^2_p + \sigma^2_e \approxeq \sigma^2_p$. 

Concerning the priors, we assume that each value within its predefined range (see Table~\ref{tab:param}) has the same probability, i.e., uniform priors.

Each spectral window contributes with the same weight to the merit function. No differential weighting scheme is applied since we have a similar number of lines for all the species.

Once the marginalized posterior probability distribution functions (PDFs) are recovered, the values of the parameters and their uncertainties are defined according to the following cases:
($a$) In the best case, when the location of the uncertainties lies within the range of possible grid-values, the solution is taken as the location of the maximum of each marginalized PDF, whilst the uncertainties are obtained as the values corresponding to the first and third quartiles of the associated cumulative distribution functions.
($b$, $c$) A lower or upper limit, when the upper or lower uncertainties lie in the upper or lower boundary limit, respectively.
($d$) An undefined solution, when the difference between the lower and upper uncertainties extends to more than 70\% of the range of possible values. 

An exception to case $c$ applies to the helium surface abundance, which, as indicated in Sect.~\ref{subsubsection:322_tmp}, has the lowest value set to \He=\,0.10. Formally then, the helium abundance is not properly determined in the analysis when its actual value is close to this value. However, the solar helium abundance limit is a reasonable ansatz for the problem at hand, and hence we adopt these cases as solutions ($a$).  

An overview of the output obtained through the spectroscopic analysis is included in Appendix~\ref{apen.output_maui} for HD\,198\,478, complemented with a  ``corner plot" to illustrate the covariance between different atmospheric parameters. We also included examples of the output distributions for each of the cases mentioned above. 


\subsubsection{Quality assessment of the solution}
\label{subsubsection:325_tmp}

\begin{figure*}[t!]
\centering
\includegraphics[width=1\textwidth]{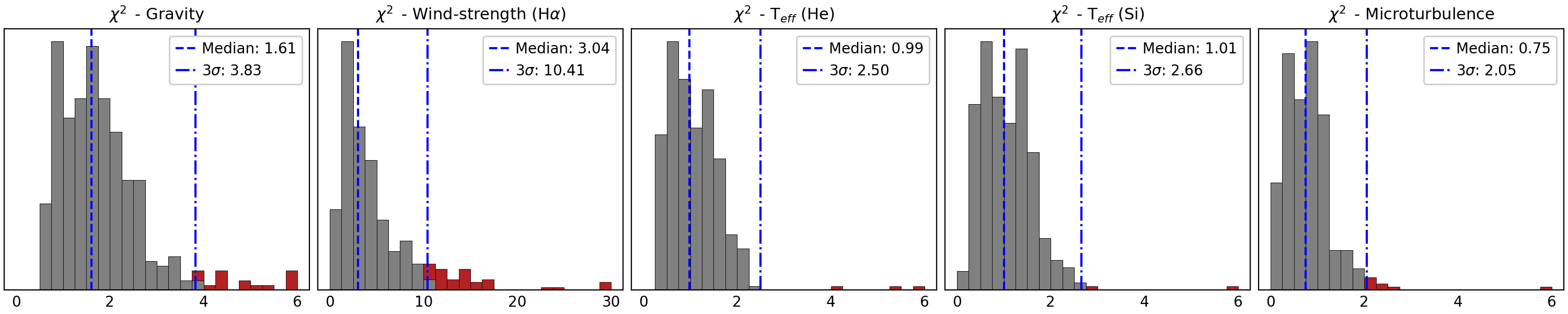}
\caption{Histograms of the five quality indicators described in Sect.~\ref{subsubsection:325_tmp}, each connected to a physical property as indicated in the title of each panel. All histograms combine the information of all the stars with reliable estimates of \Teff and \logg. The gray bins correspond to values within 3-$\sigma$ after applying an iterative clipping until convergence is achieved, while the red bins correspond to the clipped values. The median and 3-$\sigma$ values are indicated in the figure with vertical   and dashed-dotted blue lines, respectively. The associated values are shown in the legend. Note that for the panel associated with the wind-strength, the x-axis extends to significantly higher values than for the others.}
\label{fig:chi2_quality}
\end{figure*}

Given the large number of stellar spectra for which we intend to extract fundamental parameters, we decided to evaluate the quality of each solution by defining several quality indicators, each connected to some extent to one of the physical parameters: H$\alpha$ as the only indicator of the quality of the wind strength; H$\delta$, H$\gamma$, and H$\beta$ Balmer lines as indicators of the surface gravity; Si~{\sc ii-iii-iv} and He~{\sc i-ii} lines as separate indicators of the effective temperature through their ionization balances, and the Si~{\sc iii} triplet for the microturbulence, as explained in Sect.~\ref{subsubsection:323_tmp}.
The spectral window associated with each line is evaluated regarding the residuals between the observed spectra and the solution model as 
\begin{equation}
\chi_{\rm line}^2\left(O_\lambda, M_\lambda\right)=\frac{1}{n_{\rm p}} \sum_{j=1}^{n_{\rm p}}\left(\frac{O_j-M_j}{\epsilon}\right)^2,
\label{eq:chi2}
\end{equation}
where the tolerance $\epsilon$, defined as $\epsilon$\,=\,[S/N]$^{-1}$, is a measurement of how much deviation is ``allowed" between observed ($O_\lambda$) and synthetic profiles ($M_\lambda$), and $n_{\rm p}$ is the number of wavelength points in the spectral window that effectively contributed to the evaluation of the goodness-of-fit. As a result of some systematic broadening issues in some of the lines (see below), high S/N values might lead to large $\chi^{2}$ values that do not necessarily indicate a bad solution. To reduce the effect caused by these systematics, we set 100 as the upper limit of the S/N.

For each quality indicator, except the one associated with the wind, we averaged the $\chi^{2}$ values of the associated lines. In the case of Si~{\sc ii-iii-iv} and He~{\sc i-ii}, we first averaged over those lines that correspond to the same ionization stage. Finally, we used all the solutions in which \Teff and \logg correspond to case $a$ (see Sect.~\ref{subsubsection:324_tmp}) to obtain a histogram for each quality indicator (see Fig.~\ref{fig:chi2_quality} and Sect.~\ref{subsection:41_tmp}). 

We also examined the individual $\chi^{2}$ distributions of the silicon and helium diagnostic lines, which allowed us to identify systematic differences between observations and models. In particular, we found what appears to be a small but systematic difference in the broadening affecting the He~{\sc i} $\lambda$4387.93\,{\AA} and He~{\sc i} $\lambda$4921.93\,{\AA} lines. This could be related to issues with the forbidden components and not the broadening per se, and are in any case smaller than the differences found in He~{\sc i} $\lambda$6678\,{\AA}, which we considered large enough to be initially excluded from the analysis. 
We also found difficulties in reproducing He~{\sc i} $\lambda$5875.62\,{\AA} when the effect of the wind becomes relevant. 
We decided to exclude these three lines only from the following quality assessment. 

\begin{figure*}[t!]
\centering
\includegraphics[width=0.9\textwidth]{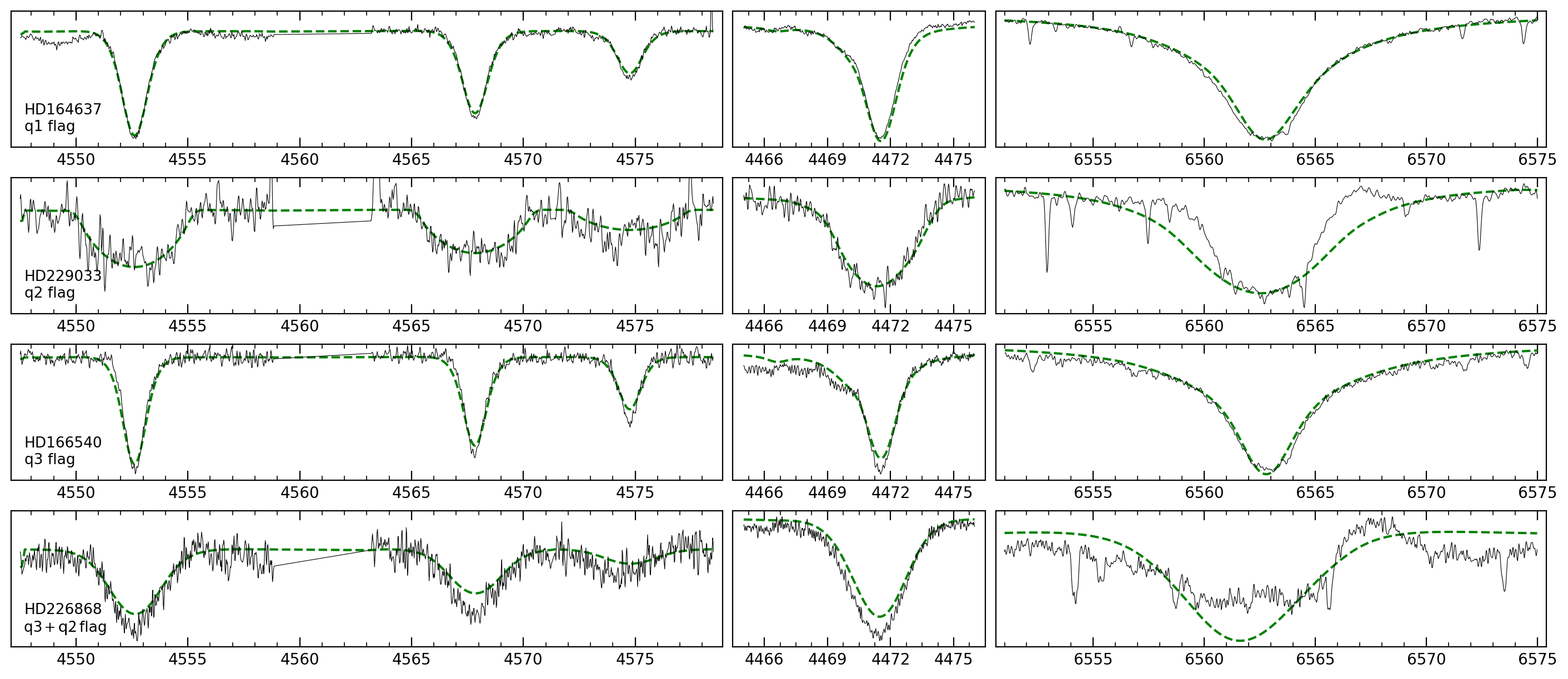}
\caption{Four illustrative cases of the quality labels assigned to the solutions (see text). Each row is divided into three spectral windows presenting three of the diagnostic regions: Si~{\sc iii} $\lambda\lambda$4553-68-75\, triplet left; He~{\sc i} $\lambda$4471\, middle; and H$\alpha$, right. The dashed green line is the synthetic spectrum of the model with the best-fitting parameters, with the solid black line being the observed spectrum.}
\label{fig:qualities}
\end{figure*}

To provide a single quality flag ($q$) that reflects the overall goodness of each solution, we used the above-mentioned histograms to consider four cases. From better to worse: 

\begin{itemize}
    \item $q1$: When each of the five values of $\chi^{2}$ lies within 3-$\sigma$ of the distribution after applying an iterative clipping of the outliers until convergence is achieved. This corresponds to a very good overall fit of all the diagnostic lines and the best reliability of the derived parameters. \smallskip
    
    \item $q2$: When the value of $\chi^{2}$ associated with $H\alpha$ lies outside 3-$\sigma$ of the corresponding distribution (see second panel of Fig.~\ref{fig:chi2_quality}), a situation which is normally indicating a non-optimal fit to the specific profile-shape of this line (see the two examples in Fig.~\ref{fig:qualities}).
    As the purpose of this work is not to provide an accurate description of the wind properties, we consider this group to be the second best if all other four values lie inside 3-$\sigma$. \smallskip
    
    \item $q3$: When one of the values of $\chi^{2}$ associated with the gravity determination, the helium or silicon ionization balances, or the microturbulence lies outside 3-$\sigma$ of the corresponding distribution (see panels 1, 3, 4, and 5 in Fig.~\ref{fig:chi2_quality}, respectively), indicating potential issues with the estimation of  \logg, \Teff, \vmic, or surface abundances. \smallskip 
    
    \item $q4$: The same as $q3$, but with two or more values of $\chi^{2}$ lying outside 3-$\sigma$. This corresponds to the worst case and is typically associated with problems in the spectrum (e.g., low S/N or normalization issues).
\end{itemize}

We note that $q3$ and $q4$ are independent of $q2$ and therefore one solution can simultaneously attributed to $q2$ and $q3$ or $q4$. Some examples of the different flags are presented in Fig.~\ref{fig:qualities}, where a comparison between observed and synthetic profiles can be found.

\section{Results}
\label{section:4_tmp}


\subsection{General outcome from the analysis}
\label{subsection:41_tmp}

Given the boundaries in \Teff and \logg of our considered grid of models, we were able to obtain estimates for a total of 527 stars (i.e. they belong to case $a$ as defined in Sect.~\ref{subsubsection:324_tmp}). The corresponding quality distribution will be detailed later below, but we anticipate already here that the majority of stars (86\%) belongs to $q1$.

The remaining 140 of the initial 666 O9\,--\,B9 type stars are not considered in the following sections and figures. They correspond to cases in which \Teff or \logg are lower or upper limits (cases $b$ and $c$, respectively), or to undefined solutions (case $d$). Concerning \Teff, we found no stars with case $b$, indicating that all O9 stars fit within the limits of the grid, but we found 57 stars with case $c$, corresponding to B6\,--\,B9 stars with \Teff values lower than the cold boundary of the considered grid of {\sc FASTWIND} models (see Table~\ref{tab:param}). Regarding \logg, 5 stars correspond to case $b$, and 34 to case $c$, the latter being mainly early-B giants and dwarfs. In 22 cases we found a combination of the previous cases. The remaining 22 stars correspond to undefined solutions.

Table~\ref{tab:errors} provides a summary of the typical formal uncertainties associated with each investigated parameter. Although our analysis provides a lower and upper error for each parameter, both were very similar (on average, less than 10\% different except for \He), and we simply considered the average values for the table. 

Figure~\ref{fig:chi2_quality} displays the histograms of $\chi^{2}$ for the five quality indicators described in Sect.~\ref{subsubsection:325_tmp}. In each panel, the position of the 3-$\sigma$ value used to assign the quality flags is included. Ideally, following Eq.~\ref{eq:chi2}, these values should gather around unity. It is evident that all the histograms except the one related to the wind-strength have median values close to unity. This indicates that the chosen 3-$\sigma$ clipping value is sufficiently stringent and that there are no significant systematic errors. In these cases, we also obtained similar 3-$\sigma$ values. However, the histogram associated with the wind-strength parameter (H$\alpha$) displays larger median and 3-$\sigma$ values, approximately three times larger. We will return to this problem in Sect.~\ref{subsection:55_tmp}.

Following the criteria described in Sect.~\ref{subsubsection:325_tmp}, we assign one of the four quality flags to each of the solutions. The percentages of solutions associated with each of them are: 83\% for $q1$, 9\% for $q2$, 7\% for $q3$, and 1\% for $q4$. Remarkably, we can see that most of them are concentrated in $q1$, indicating an overall high quality of our results. The second largest group corresponds to $q2$. This, together with the fact that there are only six $q4$ cases, and those in $q3$ correspond to spectra with low S/N or specific issues\footnote{For example, the case shown in the bottom panel of Fig.~\ref{fig:qualities}, which corresponds to the X-ray binary system HD\,226\,868 (Cyg-X1) an O9.7Iab star orbiting a black hole.}, tells us that the considered grid is suitable for the analysis of the stars that fit within the boundaries of \Teff and \logg.

The basic information about the stars in the sample is summarized in the first columns of Table~\ref{tab:qsa_results}. They include an identifiable name (ID) in the SIMBAD astronomical database \citep{weis20}, the Galactic coordinates, and the spectral classification. The following columns summarize the main outcome of the analysis, including the estimates of the rotational and macroturbulent velocities (columns \vsini and \vmac), and the estimates of \Teff, \logg, \vmic, \He and \logQ (columns are named with the abbreviations from Table~\ref{tab:param}). Except for \Teff and \logg, each of these columns is preceded by an additional column ``l", indicating which of the four possible scenarios for the probability distribution applies (see Sect.~\ref{subsubsection:324_tmp}). In particular, the upper and lower limits are indicated with ``<" and ``>", the degenerate cases with ``d", and the rest are considered reliable with ``=". An extra column named ``$q$" indicates the corresponding quality flag ($q1$-$q4$).
The last two columns indicate the name of the fits-file in the format of the IACOB spectroscopic database corresponding to the best spectrum, and the associated S/N in the 4000-5000\AA region. 

As indicated in Sect.~\ref{section:1_tmp}, metal abundances will be discussed in a forthcoming paper; however, we briefly summarize here the main outcome of our spectroscopic analysis regarding silicon abundances. Globally speaking, the associated distribution has a mean and a standard deviation of 7.46 and 0.14~dex, respectively. This result is consistent with \citet{hunter09}, who considered 56 Galactic B-stars (less than half being supergiants) located in specific clusters, obtaining $\epsilon_{Si}$\,=\,7.42\,$\pm$\,0.07~dex. Also, we find a fairly good agreement with \citet{nieva12} who obtained $\epsilon_{Si}$\,=\,7.50\,$\pm$\,0.05~dex using a sample of 20 Galactic B-type dwarfs in the Solar vicinity. Interestingly, the standard deviation of our distribution of estimated abundances is somewhat larger compared to \citet{hunter09}, \citet{nieva12}, and also compared with the typical uncertainties resulting from our analysis (see Table~\ref{tab:errors}); however, this could be related to the fact that, as shown in \citet{deburgos20}, our sample includes stars from many different locations and is certainly not limited to stars within 500\,pc from the Sun as in \citet{nieva12}, but up to 3\,--\,4\,kpc instead. Further results and discussions on these issues will be presented in a follow-up study.

\begin{table}[!t]
 \centering
\caption[]{Summary of the formal uncertainties obtained for each parameter.} 
 \label{tab:errors}
 \begin{tabular}{cccc}
 \hline
 \hline
    \noalign{\smallskip}
     Parameter & Uncertainty & Parameter & Uncertainty \\
    \noalign{\smallskip}
    \hline
    \noalign{\medskip}
    \Teff [K]       & 500$^{+200}_{-200}$ & \He  & 0.02$^{+0.01}_{-0.00}$ \\  
    \noalign{\medskip}
    \logg [dex]     & 0.07$^{+0.03}_{-0.02}$ & \vmic [km/s]   & 1.5$^{+0.7}_{-0.4}$   \\
    \noalign{\medskip}
    $-$\logQ        & 0.12$^{+0.06}_{-0.05}$ & $\epsilon_{\rm Si}$ & 0.08$^{+0.05}_{-0.02}$  \\ 
    \noalign{\medskip}
    $\beta$         & 0.41$^{+0.13}_{-0.13}$  & & \\
    \noalign{\medskip}\hline
 \end{tabular}
 \tablefoot{The positive and negative values associated with each uncertainty correspond to the third and first quartiles of the distribution, respectively.}
\end{table}


\subsection{Comparison with previous results}
\label{subsection:42_tmp}

\begin{figure*}[t!]
\centering
\includegraphics[width=1\textwidth]{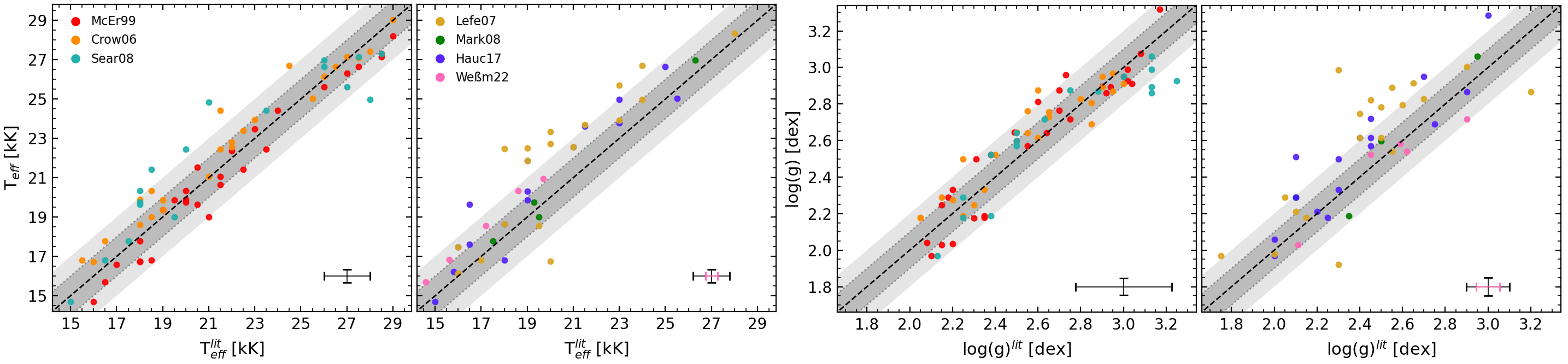}
\caption{Comparison of the results of the \Teff and \logg with previous studies in the literature. Acronyms follow those in Table~\ref{tab:lit}. The error bars in the bottom right corners indicate the average uncertainty from our analysis (vertical axis) or from the literature (horizontal axis) except those from \citet{webmayer22} for which a separate error bar in pink has been included. The two shaded areas indicate a difference in \Teff and \logg of 1000\,K and 0.1\,dex, and 2000\,K and 0.2\,dex, respectively. The diagonal black line indicates the 1-to-1 agreement.}
\label{fig:compare_lit}
\end{figure*}

\begin{table*}
  \centering
  \caption[]{Summary of the analyses used by other studies of Galactic luminous blue stars.} 
  \label{tab:lit}
  \begin{tabular}{llllllll}
   \hline
   \hline
   \noalign{\smallskip}
   Reference & Acronym & Stars in & Codes$^{a}$ & Obtain & \vmic   & Resol. & Comments$^{b}$ \\
   paper     &         & common   &             & \vmac? & [$km/s$] &        &               \\
   \noalign{\smallskip}
   \hline
   \noalign{\smallskip}
\text{\cite{mcerlean99}} &McEr99 &29 &{\tt TDS}     &No  &...        &   5000 &Unblanketed models \\
\text{\cite{crowther06}} &Crow06 &22 &{\tt CMFGEN}  &No  &10\,--\,40 &\ls5000 &Fixed \He=\,0.2 \\ 
\text{\cite{searle08}}   &Sear08 &17 &{\tt CMFGEN}  &No  &10\,--\,50 &\ls2000 &Optical + UV spectra \\
\text{\cite{lefever07}}  &Lefe07 &20 &{\tt FASTWIND}&Yes &5/10/15    &70\,000 &Fixed He and Si (solar) \\
\text{\cite{markova08}}  &Mark08 &3  &{\tt FASTWIND}&Yes &4\,--\,20  &15\,000 &Closest comparison \\ 
\text{\cite{haucke18}}   &Hauc18 &16 &{\tt FASTWIND}&Yes &5\,--\,25  &13\,000 &Fixed He and Si (solar) \\
\text{\cite{webmayer22}} &Weßm22 &5  &{\tt ADS}     &Yes &0\,--\,16  &48\,000 &Use turbulent pressure \\
   \noalign{\smallskip}
   \hline
  \end{tabular}
\begin{list}{}{}
\item {\bf Notes}. $^{(a)}$ {\tt TDS} refers to {\tt TLUSTY}, {\tt DETAIL} and {\tt SURFACE} while {\tt ADS} refers to {\tt ATLAS12}, {\tt DETAIL} and {\tt SURFACE}. $^{(b)}$ Solar He and Si corresponds to \He=\,0.1 and $\epsilon_{\rm Si}$\,=\,7.55, respectively.
\end{list}
\end{table*}

Figure~\ref{fig:compare_lit} compares our results for \Teff and \logg with other relevant studies in the literature that also performed quantitative spectroscopic analysis on small/medium-sized samples of Galactic luminous blue stars. They are separated into two groups to better illustrate the differences. Table~\ref{tab:lit} summarizes the main characteristics of the different analyses used by those works and the number of stars in common. The table also includes the different acronyms used to refer to each of those works. In most cases, they have made use of the atmospheric codes {\tt CMFGEN} \citep{hillier98}, or {\tt FASTWIND} as also done here. Other cases include the use of {\tt TLUSTY} \citep{hubeny88} or {\tt ATLAS12} \citep{kurucz05} combined with {\tt DETAIL} \citep{giddings81} and {\tt SURFACE} \citep{butler85}. In \citet{mcerlean99}, \citet{crowther06}, and \citet{searle08} (first and third panels of Fig.~\ref{fig:compare_lit}), the typical formal uncertainties in \Teff and \logg are $\approx$2000\,K and $\approx$0.2\,dex, respectively. In \citet{lefever07}, \citet{markova08}, and \citet{haucke18} (second and fourth panels), the uncertainties are on average $\approx$800\,K and $\approx$0.1\,dex, respectively. \citet{webmayer22} claims the smallest average uncertainties of $\approx$250\,K and $\approx$0.05\,dex. We note that, except for \citet{webmayer22}, the quoted uncertainties are systematically larger than those reached in our analysis (see Table~\ref{tab:errors}); this is mostly a consequence of our analysis method and the way our uncertainties are derived.

The different strategies and methodologies used in those works make it very difficult to assess the overall agreement with our results, as well as to carry out individual comparisons. Despite this, we provide some individual notes and try to explain the reasons for some of the more notorious differences. 

First, we find a good overall agreement with the results of \citet{mcerlean99}, one of the first studies attempting to derive spectroscopic parameters for a large sample of BSGs. Most of the results lie within $\pm$1000\,K and $\pm$0.1\,dex as shown in the corresponding panels of Fig.~\ref{fig:compare_lit}. This is interesting as it is a study with large differences in the methodology, as they used plane-parallel geometry and unblanketed models. 

The comparison with \citet{crowther06} also shows a very similar situation. However, one main difference from this work is their use of a fixed \He=\,0.2 which, as shown in Sect.~\ref{subsection:54_tmp}), is not a representative value for the majority of the analyzed luminous blue stars. Comparing the fit quality for those stars with differences in \Teff or \logg larger than 1000\,K or 0.1\,dex, respectively, we find a typically better quality from our results.

In the case of \citet{searle08}, we observe a larger scatter of the differences, both in \Teff and \logg. The latter was also found by the authors themselves when comparing with \citet{crowther06}, being of the order of 0.1\,--\,0.2 dex. Despite that they attributed this difference to ``wind contamination" of the Balmer lines, one also finds differences in some diagnostic metallic lines and significant discrepancies between their fitted and observed spectra.

The differences with \citet{mcerlean99}, \citet{crowther06}, and especially \citet{searle08} might also be attributed to the much lower resolutions used in those works compared to our data. Moreover, none of these works accounted for macroturbulent broadening, which can represent an important contribution to the shapes of the lines.

Our comparison with \citet{lefever07} shows the largest discrepancies with our results, with lower \Teff and \logg values for many of the stars in common. One reason for this difference could be their (much) lower number of diagnostic lines compared to other studies. In particular, they only used H$\gamma$ as the primary gravity indicator and He~{\sc i} $\lambda$4471.47\,{\AA} in the second place. For \Teff they used either the Si~{\sc ii} 4130\,{\AA} doublet or the Si~{\sc iii} 4560\,{\AA} triplet. They also adopted a fixed solar silicon abundance, which can also affect the determination of the effective temperature.

The study by \citet{markova08} represents the closest comparison to our methodology. However, the number of stars in common is very limited. Nevertheless, we observe a good agreement for almost all stars in common.

The results from \citet{haucke18} show a good agreement for half of the stars in common, with the other half having lower \Teff and \logg values than in our case. For these stars, we found (as also by the authors) that their \Teff and \logg values are systematically lower compared to other studies such as \citet{crowther06} or \citet{searle08}. 

Last, our results compared to \citet{webmayer22} show a good agreement despite the different methodologies and the fact that they account for the effects of turbulent pressure on the models. We could only identify a slight trend towards lower \Teff in their case. The authors suggested (via private communication) that the differences are likely due to differences in the silicon model, especially affecting some Si~{\sc ii} lines \citep[see][for more details]{webmayer22}.

In summary, we do not see any particular trend in our results that may indicate a problem with our models or with the analysis technique. We also do not find particular differences when comparing the results obtained with {\tt FASTWIND} or {\tt CMFGEN}. Regarding the largest differences, they were attributed in the first place to specific reasons related to the fit quality \citep[e.g.][]{searle08}, or the absence of diagnostic lines \citep[e.g.][]{lefever07}.


\subsection{\texorpdfstring{\Teff--\,Spectral type calibration}{Teff -- spectral type calibration}}
\label{subsection:43_tmp}

\begin{figure}[!t]
\centering
\resizebox{\columnwidth}{!}{\includegraphics{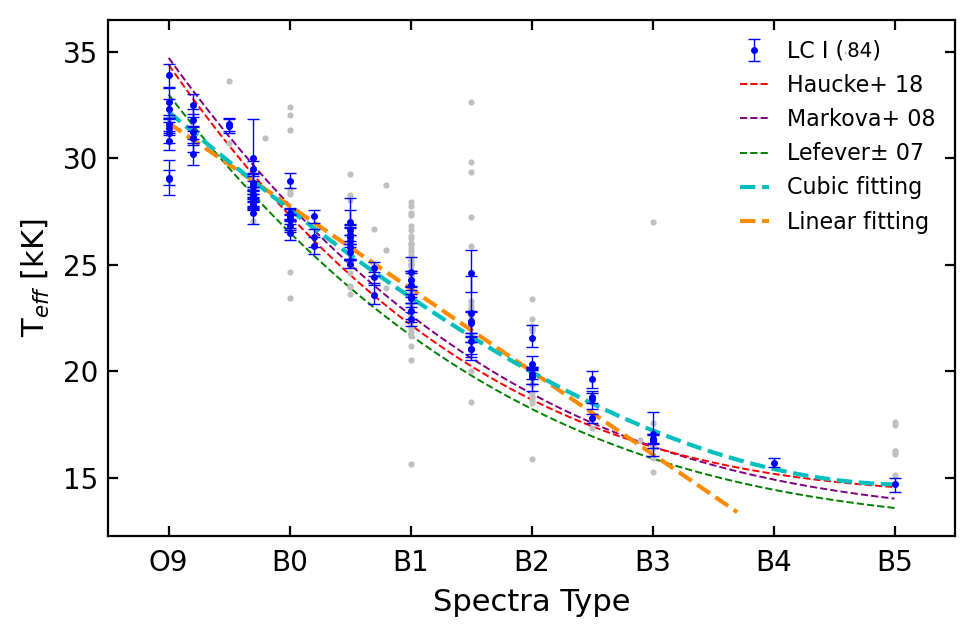}}
\caption{\Teff against SpT for stars with LC\,I. Blue dots correspond to stars with revised classifications (see Sect.~\ref{subsection:43_tmp}), whereas the gray dots correspond to stars whose classification corresponds to the default one provided by SIMBAD. The dashed orange and cyan lines correspond to a first and third-order polynomial fit to stars colored in blue. Some previous calibrations from the literature are also included for comparison.}
\label{fig:calib_spt}
\end{figure}

Several studies have obtained calibrations of spectral type (SpT) against \Teff for Galactic BSGs in the past \citep[see][]{lefever07, markova08, searle08, haucke18}. These calibrations are, however, based on samples of relatively small size, with no more than a few tens of targets in the best cases. Here, we benefit from our much larger sample to provide a revised calibration.
Figure~\ref{fig:calib_spt} shows \Teff as a function of SpT for those analyzed stars with luminosity classes Ia, Iab, and Ib. To avoid spurious results associated with the use of erroneous spectral classifications \citep[as those provided by SIMBAD in many cases, see][and Sect.~\ref{subsection:44_tmp}]{deburgos23}, we highlight with blue dot symbols those stars with reliable spectral classifications as provided by \citet{sota11}, \citet{sota14}, \citet{deburgos20, deburgos23}, and \citet{negueruela-subm}.

Using only those stars in the former group, we performed both a linear and a third-order polynomial fit \citep[the later option first proposed by][]{lefever07}, accounting for individual errors in \Teff.
The resulting calibrations are:
\begin{align*}
\Teff &= 27719 - 3754x\,\,[\rm K]\\
\Teff &= 27597 - 4104x + 130x^2 + 33x^3\,\,[\rm K]
\end{align*}
where $x$ is the SpT adopting O9\,=\,-1, B0\,=\,0, and so on. Additionally, the 1-$\sigma$ uncertainties for each SpT are: O9\,$\pm$\,1300\,K, B0\,$\pm$\,1500\,K, B1\,$\pm$\,700\,K, B2\,$\pm$\,700\,K, B3\,$\pm$\,600\,K, where for the B4- and B5-type stars, they are not provided due to the reduced number of objects.

As illustrated in Fig.~\ref{fig:calib_spt}, both calibrations -- indicated with black and gray dashed lines, respectively -- are almost identical from O9 down to B3-type stars, but significantly differ for later types. We also see a much larger scatter in \Teff for those stars whose classifications are directly extracted from SIMBAD (gray dot symbols).

Compared to previous calibrations by \citet{lefever07}, \citet{markova08}, and \citet{haucke18}, their regression curves seem to agree with our results only for B0-type objects, differing by 1\,--\,3\,kK for O9 and B1\,--\,B5 spectral types. The explanation for this difference is the considerably smaller number of targets considered by these authors, together with the change of slope beyond the B5 spectral types, which notably modify the polynomial fits. Taking into account the improved statistics, our calibration is clearly more robust than the other three.

\subsection{Spectroscopic HR diagram}
\label{subsection:44_tmp}

\begin{figure*}[t!]
\centering
\includegraphics[width=0.9\textwidth]{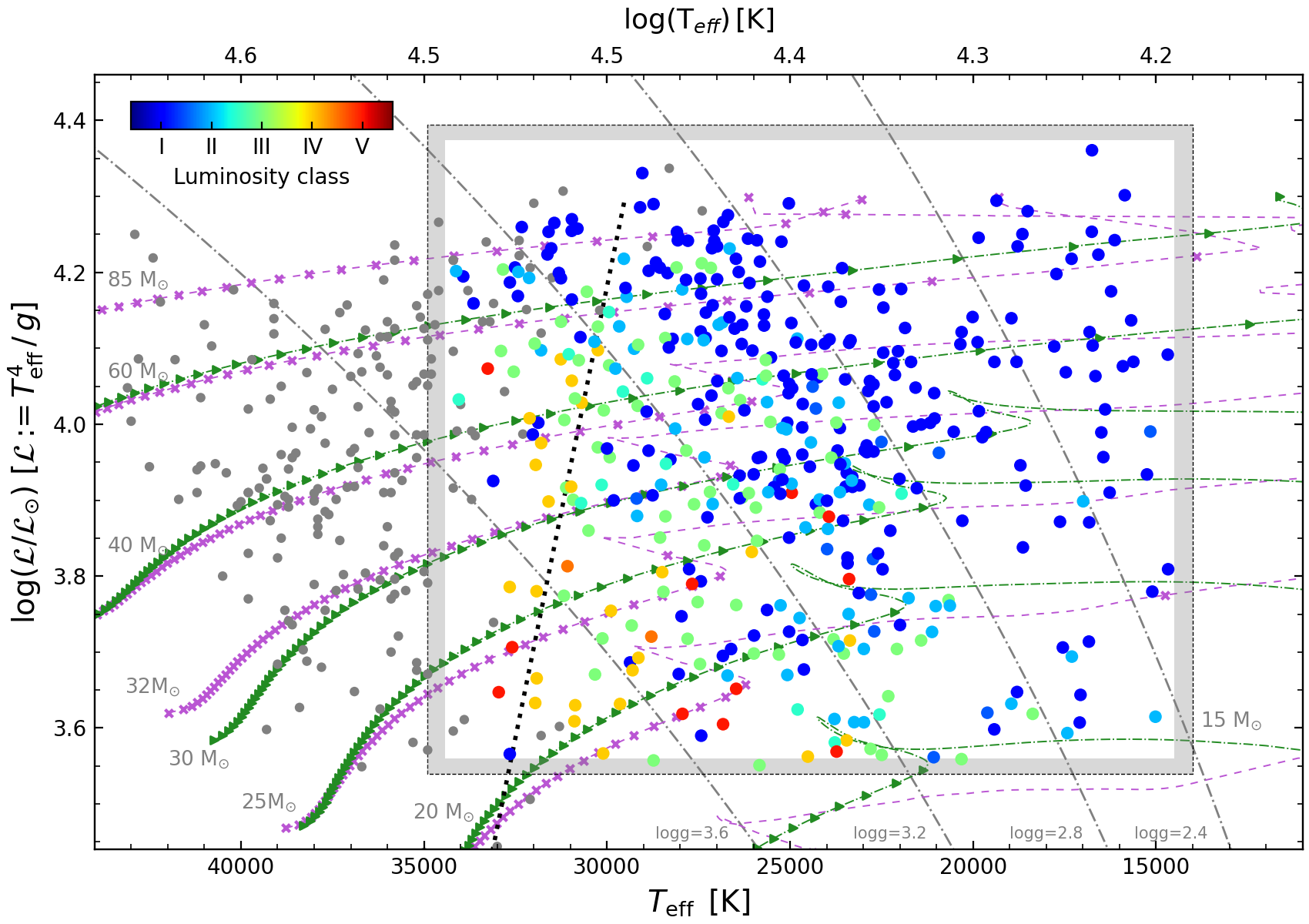}
\caption{sHR diagram showing our results from the analysis for 527 stars with O9\,--\,B5 spectral type color-coded by their luminosity class, and 191 O-type stars from Hol18-22 in gray. 
The boundaries of our model grid are indicated with a rectangle. The shady area indicates the approximate region where our results correspond to the upper or lower limits (see Sect.~\ref{subsubsection:324_tmp}). The approximate separation between the O- and B-type stars is indicated with a dotted diagonal black line. 
For reference, the figure includes non-rotating evolutionary tracks with solar metallicity from the Geneva and Bonn models (\citealp{ekstrom12, georgy13} and \citealp{brott11}, respectively). Intervals of the same age difference are marked with purple crosses for Geneva and green triangles for Bonn, which are connected with dashed and dashed-dotted lines of the same color, respectively. The dashed-dotted gray lines indicate different constant \logg values.}
\label{fig:shrd}
\end{figure*}

The location of the sample stars in the spectroscopic Hertzsprung–Russell diagram \citep[hereafter sHR diagram][]{langer14} is shown in Fig.~\ref{fig:shrd}, where we also indicate the boundaries of our model grid. Along with the stars in our study, the figure also includes 191 O-type stars from \citet{holgado18, holgado20, holgado22} (hereafter Hol18-22).

The colors in Fig.~\ref{fig:shrd} indicate the luminosity class of the stars as listed in Table~\ref{tab:qsa_results}. In particular, for most of the stars in the sample, we adopted the recommended classifications quoted in SIMBAD. However, as shown in \citet{deburgos23}, for B-type stars, a non-negligible number of LCs in SIMBAD are incorrect or not even provided (see also Fig.~\ref{fig:calib_spt}). While we plan to review the spectral classifications of all B-type stars in our sample following the guidelines of \citet{negueruela-subm}, for this work we keep using the SIMBAD classifications except for those $\approx$120 stars for which we have published revised spectral types and luminosity classes \citep[see][]{deburgos20, deburgos23, negueruela-subm}. These revised classifications represent an improvement for supergiant stars \citep[see for comparison Fig.~5 in][]{deburgos23}.

As illustrated in Fig.~\ref{fig:shrd} (see also Sect.~\ref{section:2_tmp}), the majority of stars in our sample ($\approx$70\%) comprise stars with LC\,I and II, especially toward cooler temperatures (below \Teff\ls\,29\,kK, see also Fig.~\ref{fig:hist_teff}). However, there is also a non-negligible number of LC\,III, IV and V objects. Despite most of them being located close to the hot boundary of the investigated domain, there is still a fraction of objects from this latter group whose location in the sHR diagram overlaps with the region predominantly populated by LC\,I\,--\,II stars. Following the criteria formulated in \citet{negueruela-subm}, we revised the spectral classification for those LC\,V stars that overlap with the location of stars with LC\,I, II, and III. Appendix~\ref{apen.dwarfs} shows that almost all of them actually correspond to stars with LC\,III.

This result warns us again about the use of unchecked spectral classifications from SIMBAD and highlights the urgent need for a systematic revision of an important percentage of the known B-type stars, following a similar homogeneous approach as the work performed by \citet{maiz-apellaniz11, maiz-apellaniz16, sota11, sota14} in the case of O-type stars.


\section{Discussion}
\label{section:5_tmp}


\subsection{An empirical hint for the Terminal Age Main Sequence in the high mass domain?}
\label{subsection:51_tmp}

The empirical identification of the location of the Terminal Age Main Sequence (TAMS) provides important constraints for several physical phenomena occurring in the interior of stars along the Main Sequence (MS), including core overshooting processes, and the impact of rotational mixing and magnetic fields, among others \citep[see, e.g.,][]{meynet00, vink00, maeder05, schootemeijer19, martinet21, scott21}. Above $\approx$3\MSol, once hydrogen is exhausted in the convective core and the TAMS is reached, stars suffer from a rapid reconfiguration of their internal structure while evolving approximately at constant luminosity. In brief, they increase considerably their size (and hence the effective temperature decreases), while the inert core is contracting. As a consequence of the short time-scale of this process, the relative number of stars in a volume-limited sample detected on the cool side of the TAMS is expected to be considerably lower than those populating the MS.

\begin{figure}[!t]
\centering
\resizebox{\columnwidth}{!}{\includegraphics{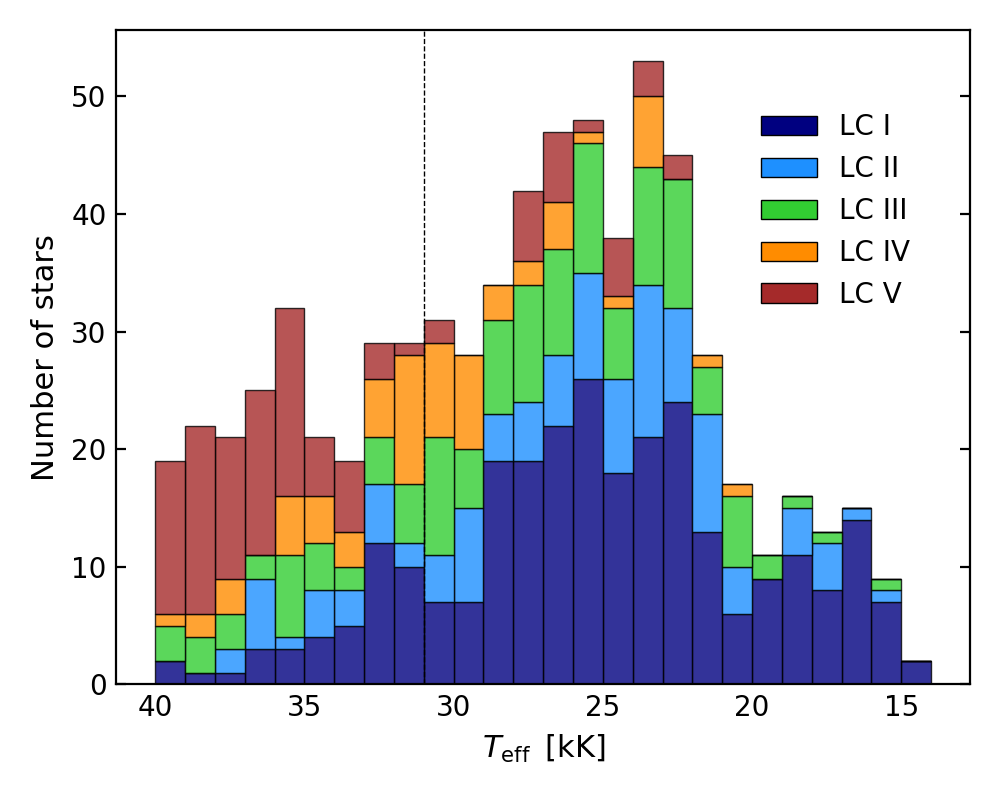}}
\caption{Histogram of effective temperatures for all the stars in Fig.~\ref{fig:shrd}, color-coded by luminosity class. The approximate separation between O and B-type stars is indicated with a dashed vertical black line.}
\label{fig:hist_teff}
\end{figure}

Figure~\ref{fig:hist_teff} depicts a histogram of the effective temperatures of the stars shown in Fig.~\ref{fig:shrd}. The relative number of stars in each bin steadily increases from the hot end down to $\approx$21\,kK, where a clearly noticeable drop is detected. As indicated above, this drop might roughly delineate the location of the empirical TAMS in the mass range between 15 and 85\MSol. Interestingly, this severe drop in the number of stars is located 5\,--\,7\,kK below the theoretical TAMS predicted by the single-star evolutionary models of \citet{ekstrom12}. Alternatively, if compared to the single-star models of \citet{brott11}, the location of the drop overlaps well with the theoretical TAMS for masses below 40\MSol, but for masses above, the TAMS is shifted to temperatures $\approx$10\,kK cooler. The large difference between both sets of models is mainly related to the different treatment of angular momentum transport (\citealp[][advective]{ekstrom12}; \citealp[][diffusive]{brott11}), and the different size of the core-overshoot parameter \citep[lower in][]{ekstrom12}, where the latter has a large impact on the MS-lifetimes.

While the presence of BSGs beyond the theoretically predicted MS in single-star evolutionary models has been known for a while \citep[see, for example,][using photometric and spectroscopic observations, respectively]{fitzpatrick90, castro14}, our work implies a higher statistical significance, given the large sample of stars homogeneously analyzed here. Should this location of the TAMS be confirmed, not all BSGs would be He-core burning post-MS stars, with a possible significant fraction of them (mostly those with spectral types earlier than B3) being H-core burning objects \citep[see also previous hints by][]{vink10, brott11, castro14, mcevoy15}.

However, the possibility of other evolutionary channels populating this part of the sHR diagram -- mainly invoking post-mass transfer binaries and mergers \citep[see][and references therein]{marchant23}, but also post-red supergiant stages through blue loops \citep[e.g.][]{stothers75, martinet21, zhao23} --, complicates a definitive identification of BSGs as MS or post-MS objects just accounting from the simple picture described above. In addition, the potential impact of observational biases, as well as sampling effects related to the initial mass function and the age range of the compiled sample should be taken into account in any attempt to explain the observed distribution of stars presented in Fig.~\ref{fig:shrd}.

For example, despite one would expect a more or less constant distribution of stars as a function of effective temperature along the MS evolution, the relative number of stars in the \Teff range $\approx$30\,--\,20\,kK is noticeably larger than in the $\approx$40\,--\,30\,kK range (see Fig.~\ref{fig:hist_teff}). This could be partially explained by the effect of a Malmquist bias affecting our magnitude-limited sample \citep[see][for a detailed discussion]{deburgos23}. Since mid B-type supergiants are expected to be intrinsically brighter in the optical than other supergiants with similar luminosities but earlier spectral types, the distances reached for the former group are much larger than for the latter; hence, an overabundance of mid-to-late BSGs is expected in our sample. This strengthens our suggestion that the TAMS might be located at $\approx$21\,kK if we assume the drop in density of stars as a function of \Teff as an empirical evidence of the position of the TAMS.


\subsection{Rotational properties}
\label{subsection:52_tmp}

\begin{figure}[!t]
\centering
\resizebox{\columnwidth}{!}{\includegraphics{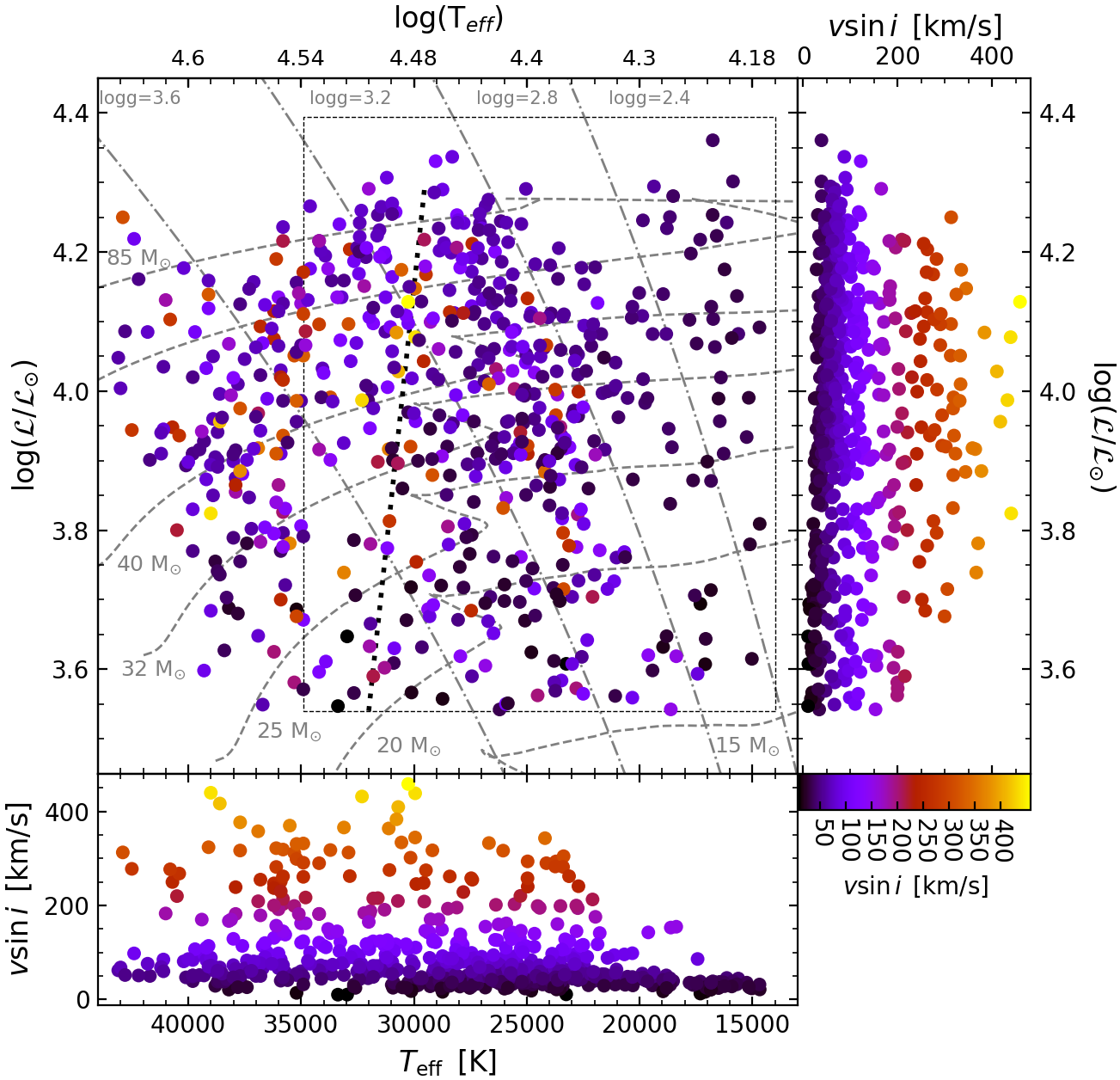}}
\caption{sHR diagram of the 527 stars in the sample and 191 O-type stars from Hol18-22, all color-coded by \vsini. The bottom and right sub-panels show \vsini against \Teff and log\,\Ls, respectively. The boundary limits of our grid of models are marked with dashed black lines. Evolutionary tracks and \logg isocontours are the same as in Fig.~\ref{fig:shrd}.}
\label{fig:shrd_vsini}
\end{figure}

\begin{figure}[!t]
\centering
\resizebox{\columnwidth}{!}{\includegraphics{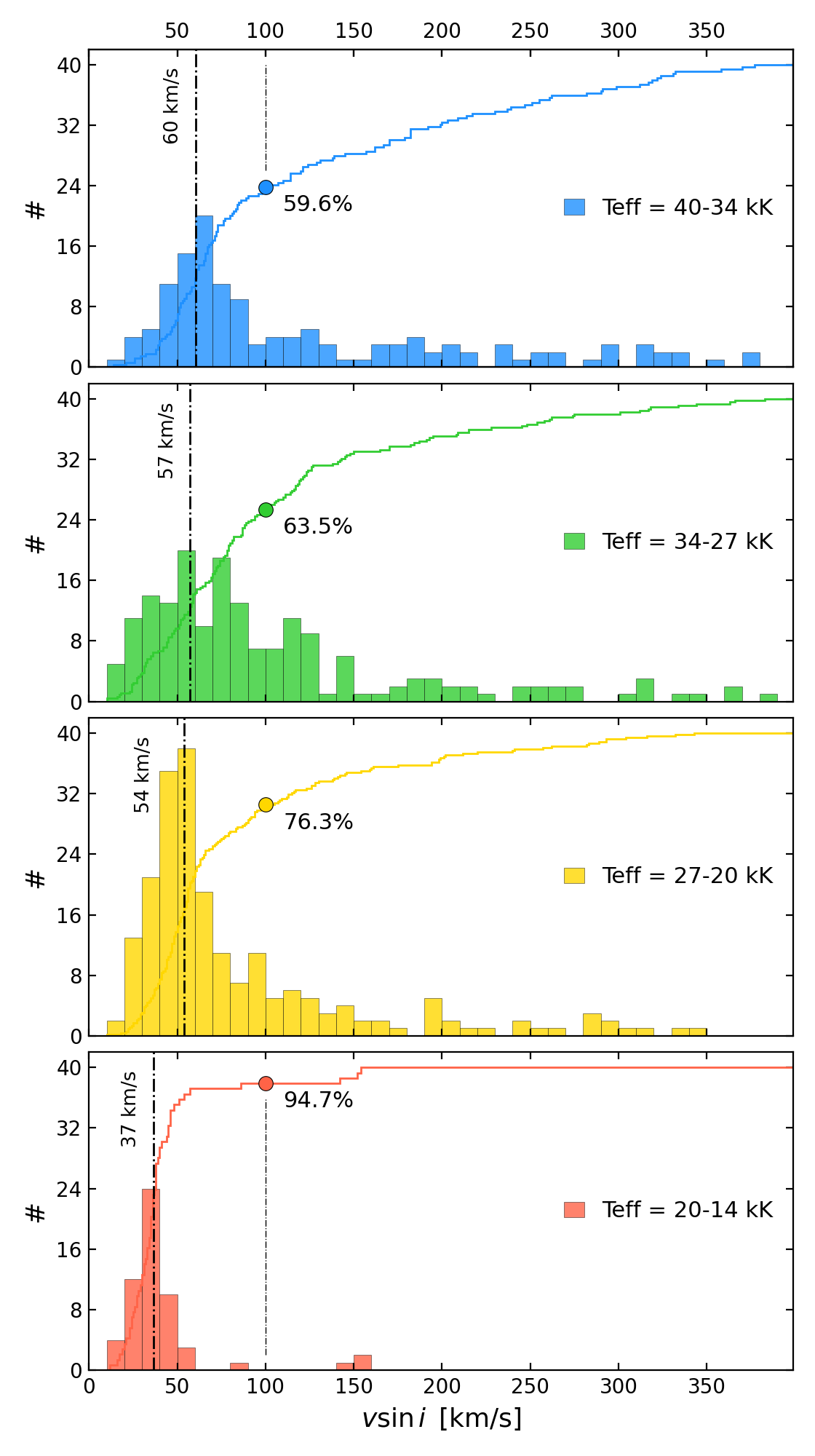}}
\caption{Distribution of \vsini separating the stars in the sample in four groups of different \Teff ranges, as shown in the legends. The panels show the different histograms, indicating with a dashed-dotted line the mean \vsini values as derived from an iterative 2-$\sigma$ clipping. In each panel, the corresponding cumulative distribution and the percentage of stars with \vsini<\,100\kms is also included.}
\label{fig:hist_vsini}
\end{figure}

Figure~\ref{fig:shrd_vsini} shows an sHR diagram with the same stars as in Fig.~\ref{fig:shrd}, but color-coded by their projected rotational velocities. The central panel is complemented with another two (right and bottom sub-panels) in which the measured \vsini values are directly confronted against log\,\Ls and \Teff, respectively.

As observed for Galactic O-type stars \citep[see][and references therein]{holgado22}, two main components can also be clearly distinguished in the \vsini distribution when moving to the BSG domain (see also Fig.~\ref{fig:hist_vsini}): one main component -- comprising about 70\% of the sample -- with projected rotational velocities ranging from $\approx$10\kms to $\approx$100\kms, and a tail of fast rotating stars reaching values of $\approx$400\kms. While the main component is present in the full range of covered effective temperatures, the tail of fast rotators disappears below $\approx$20\,kK (see bottom panel of Fig.~\ref{fig:shrd_vsini}).

The existence of a bimodal \vsini distribution in the O star domain has been known for several decades \citep[see, e.g.,][]{conti77}. This is also the case for the clear drop found in the \Teff\,--\,\vsini diagram at \Teff$\approx$\,21\,kK (see bottom panel of Fig.~\ref{fig:shrd_vsini}), which has previously been identified by several authors \citep[see, e.g.,][]{howarth97, vink10, fraser10, brott11}, including our previous work \citep{deburgos23}, where we found its location around the B2-type stars with LC\,I\,--\,II.

A theoretical explanation for the occurrence of a bimodal distribution \citep[proposed by][]{demink13} invokes the effect of mass transfer in binary systems, implying the spin-up of the gainer. In this scenario -- which has found empirical support by \citet{holgado22} and \citet{britavskiy23} -- the tail of fast rotators is mostly populated by post-interaction binary products, and the observed \vsini distribution is not necessarily representative of the initial spin-rate at birth of the investigated samples. This hypothesis leaves room for the possibility that the low \vsini component of the distribution mostly comprises stars which have not interacted with any companion, while also including some fast rotating stars seen with a low inclination angle, as well as potential mergers spun-down by magnetic fields \citep[see, e.g.,][]{schneider16, keszthelyi19}.

Thanks to the large sample of stars for which we have obtained \vsini, \Teff, log\,\Ls, and \logQ estimates, and as a follow-up of the work started in \cite{holgado22}, we can evaluate with good statistical significance and robustness how the observed \vsini distribution is modified as stars evolve. To this end, we use \Teff as a proxy of evolution, but also take into account that in the binary channel the direct relation between the \Teff and age breaks down.

Figure~\ref{fig:hist_vsini} depicts the histograms of \vsini for four sub-samples of stars covering, from top to bottom, decreasing ranges of \Teff. In particular, we consider three sub-samples covering the region between our hotter \Teff boundary and the speculated location of the TAMS (see Sect.~\ref{subsection:51_tmp}), plus a fourth one comprising the supposedly post-MS region. In all cases, we mark the location of the mean \vsini associated with the low \vsini component of the distribution and indicate the percentage of stars that have a \vsini below 100\kms.

Regarding the low \vsini component, both Fig.~\ref{fig:hist_vsini} and the bottom sub-panel of Fig.~\ref{fig:shrd_vsini} show a slow decrease of its characteristic \vsini (from 60 down to 54\kms in the \Teff range between 40 and 21\,kK, and from this later value down to 37\kms when considering the cooler stars in the sample). This result is consistent with recent findings by \citet{holgado22} for the case of O-type stars, but also extending them further to lower effective temperatures. Despite the widely predicted loss of angular momentum due to stellar winds, the detected surface braking in the low \vsini component is almost negligible throughout the considered range of effective temperatures. As suggested by \citet{holgado22}, this might be pointing towards the existence of an efficient mechanism transporting angular momentum from the stellar core to the surface along the main sequence. Indeed, this statement might also be supported by the almost constant percentage of stars with \vsini>\,100\kms, as well as the maximum \vsini values detected in the tail of fast rotators in stars ranging from the Zero-Age-Main-Sequence (ZAMS) to the suggested location of the TAMS \citep[see][and Sect.~\ref{subsection:51_tmp}]{holgado22}.

Regarding the drop in \vsini at \Teff$\approx$\,21\,kK, as pointed out by \citet{vink10}, it could be either an indicator of the end of the MS or the result of an enhanced angular momentum loss at the theoretically predicted bi-stability jump \citep{pauldrach90, vink99, vink00}\footnote{The proposed location of the bi-stability jump in \citet{vink00} for the range of luminosities of the considered BSGs is $\approx$25\,kK.} When considering this latter possibility, we must remember that the exact location and characteristics of the bi-stability jump remain a debated question \citep[see][]{petrov16, krticka24}. Indeed, there is not even consensus on the predicted occurrence of a significant increase in the mass loss rate when the star is crossing from the hotter to the cooler side of the bi-stability jump \citep{bjorklund21, bjorklund23}. Furthermore, as described in Sect.~\ref{subsection:55_tmp}, the behavior of our measured wind-strength parameter does not support a strong change of the mass loss rate properties around the effective temperature where the drop in \vsini is detected. Thus, we are still left with the question of what causes the observed drop in the \vsini distribution as a function of \Teff.


\subsection{Microturbulence and macroturbulence}
\label{subsection:53_tmp}

\begin{figure}[!t]
\centering
\resizebox{\columnwidth}{!}{\includegraphics{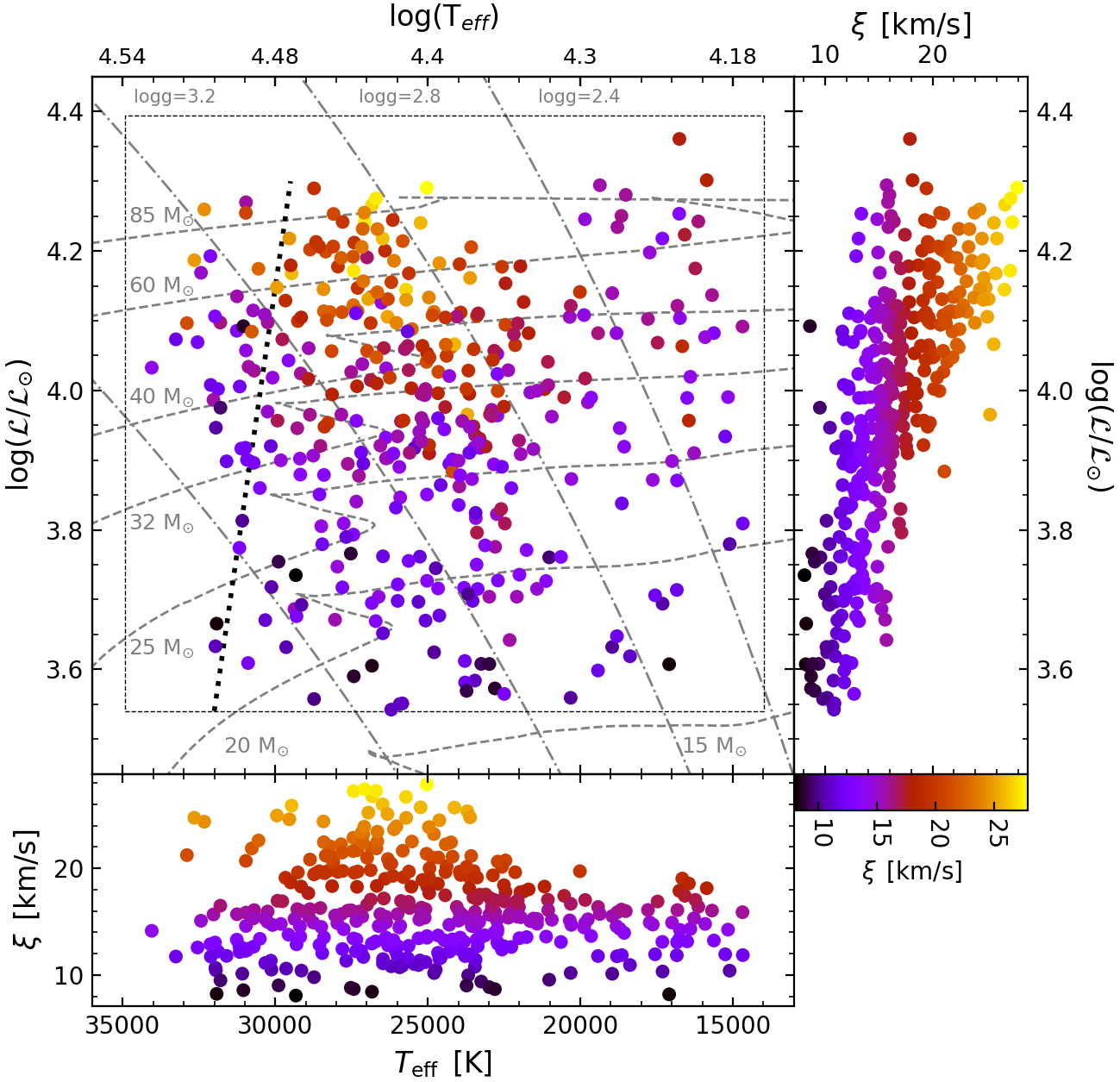}}
\caption{sHR diagram of the stars in the sample color-coded by the microturbulence. The bottom and right sub-panels show \vmic against \Teff and log\,\Ls, respectively. Results of \vmic considered as upper or lower limits, or degenerated are excluded. Evolutionary tracks and \logg isocontours are the same as in Fig.~\ref{fig:shrd}.}
\label{fig:shrd_micro}
\end{figure}  

Current analyses of the atmospheres of O- and B-type stars require the consideration of two broadening parameters, termed microturbulence \citep[see, e.g.,][]{mcerlean98, smith98, vink00} and macroturbulence \citep[e.g.][]{ryans02, simon-diaz14a, simon-diaz17}, for which their exact physical origin is yet unknown.

Figure~\ref{fig:shrd_micro} presents our derived microturbulences. Previous studies based on smaller numbers of stars in the B-stars domain have shown that supergiants have larger values of microturbulent velocities (\vmic) than giants and dwarfs \citep{gies92, hunter07, lefever07, markova08, hunter08, webmayer22}. Benefiting from a much larger sample of stars, we investigated whether a connection between the spectroscopic luminosity and the microturbulence is statistically sound. A Spearman's rank-order correlation of our results delivers a coefficient of $\rho$\,=\,0.82 with a significance level of 95\%, indicating the hypothesis of statistical independence between both quantities can be rejected.

Concerning the variation of \vmic with respect to \Teff for luminous blue stars \citep[see, for example][]{markova08}, we obtained a median value of 20\kms for O9\,--\,B0.5, 17\kms in the B0.5\,--\,B2 range, 15\kms at B2\,--\,B4 type, and 12\kms for B5 and later. We also notice an increased relative and absolute scatter towards the hotter \Teff end, being particularly broad at \Teff$\approx$\,26\,kK, whereas a smaller scatter is present at the cool end.

\begin{figure}[!t]
\centering
\resizebox{\columnwidth}{!}{\includegraphics{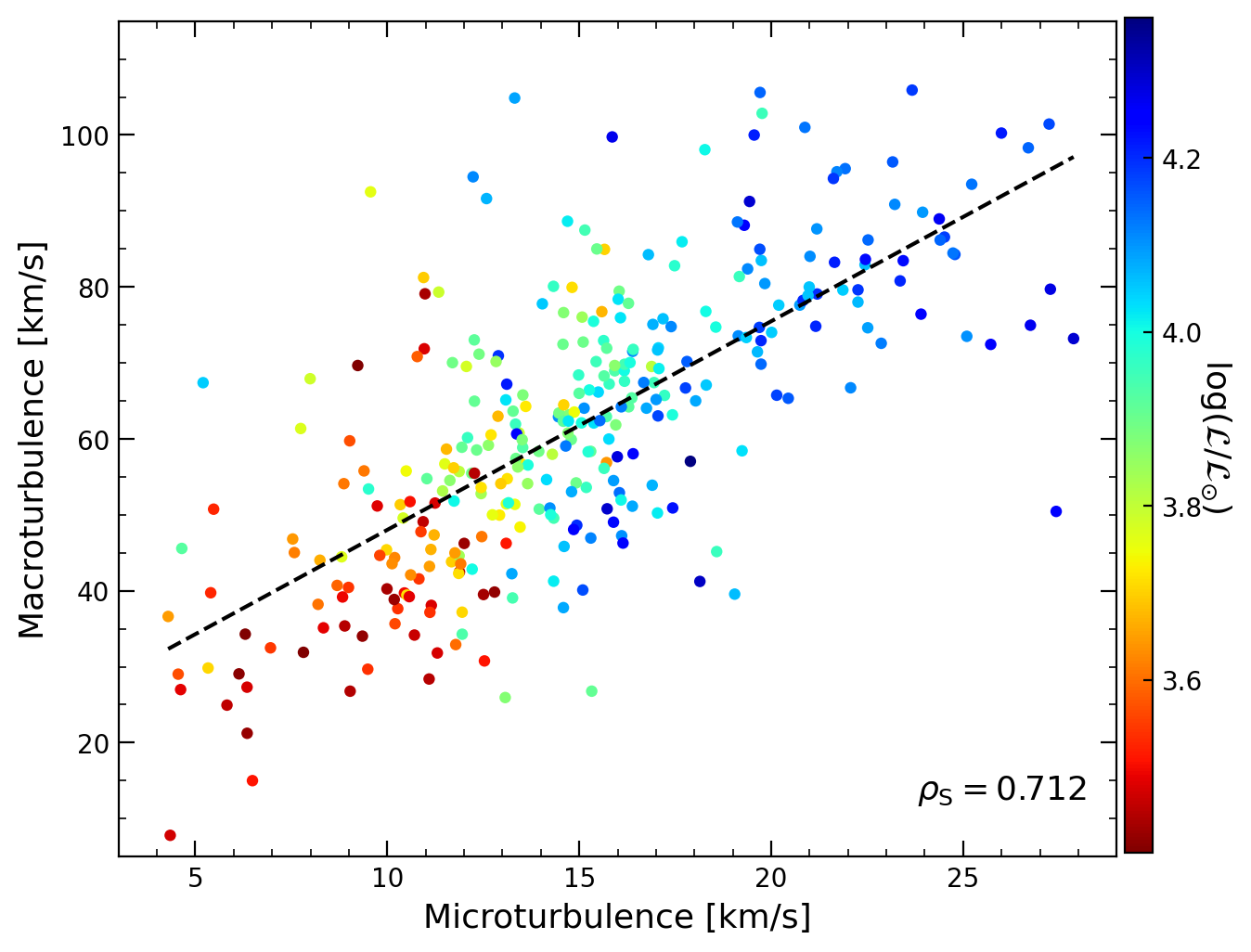}}
\caption{Macroturbulence against microturbulence for the sample of stars, color-coded by their \logLs. The sample is limited to those stars with \vsini<\,100\kms. A linear fit is included and indicated by a dashed diagonal black line.} 
\label{fig:micro_vs_macro}
\end{figure}

Our derived macroturbulent velocities (\vmac), combined with those from Hol18-22 for O-type stars, essentially reproduce the previous findings by \citet{simon-diaz17} in terms of the dependencies with respect \Teff and log\,\Ls, and hence are not repeated here. However, we go beyond that study in terms of investigating a potential correlation between \vmic and \vmac. Interestingly, Fig.~\ref{fig:micro_vs_macro} shows that a positive correlation does exist. 
To quantify this correlation we calculated Spearman's correlation coefficient, which resulted in $\rho$\,=\,0.71 at a significance level of 95\%. This result deserves a follow-up, more in-depth study since it might indicate a connection between the physical drivers of both broadening mechanisms.

Several plausible scenarios have been proposed to explain the occurrence of these two spectral line-broadening features. Among them, \citet{cantiello09} suggests that microturbulence originates in sub-surface convective zones, whereas \citet{grassitelli15a} and \citet{cantiello21} propose the same origin for macroturbulence. As an alternative, \citet{aerts09} suggested the collective pulsational velocity broadening due to gravity modes as a physical explanation for the macroturbulent broadening in hot massive stars. More recently, \citet{aerts15} extended further the proposed connection between macroturbulence and stellar variability phenomena by linking both through the effect of convectively driven waves originating in the stellar core \citep[see also][]{edelmann19, lecoanet23, anders23}. This latter scenario has been further explored by \citet{bowman19a, bowman19b, bowman20b}, who also showed evidence of a correlation between the amplitude of observed stochastic low-frequency photometric variability, and the amount of measured macroturbulent broadening. 

Overall, despite the various alternatives proposed, no firm conclusions have been reached yet \citep[see, e.g.][]{simon-diaz17, godart17, bowman20b, cantiello21}. In these regards, the empirical correlations presented here and in \citet{simon-diaz17}, together with results from parallel works investigating the connection between macroturbulent broadening and photometric and line-profile variability \citep[e.g.][]{simon-diaz10b, simon-diaz17, bowman20b} open new avenues to find more conclusive answers about the physical origin and potential connection between these two ubiquitous features.

\subsection{Surface helium abundance}
\label{subsection:54_tmp}

Together with nitrogen and carbon, a consistent determination of the surface abundances of helium in O- and B-type stars can help to constrain the impact of internal mixing processes along the main sequence evolution \citep[see, e.g.,][]{martins05, rivero-gonzalez12, carneiro16, grin17}, identify the occurrence of mass transfer and merger events in massive binaries \citep[see, e.g.,][]{langer12, langer20, demink13, glebbeek13, schneider16, sen22, menon23} and, ultimately, better identify the evolutionary status of the investigated targets \citep[e.g., whether they are in a H- or He-core burning stage; see,][and references therein]{georgy21}. 

\begin{figure}[!t]
\centering
\resizebox{\columnwidth}{!}{\includegraphics{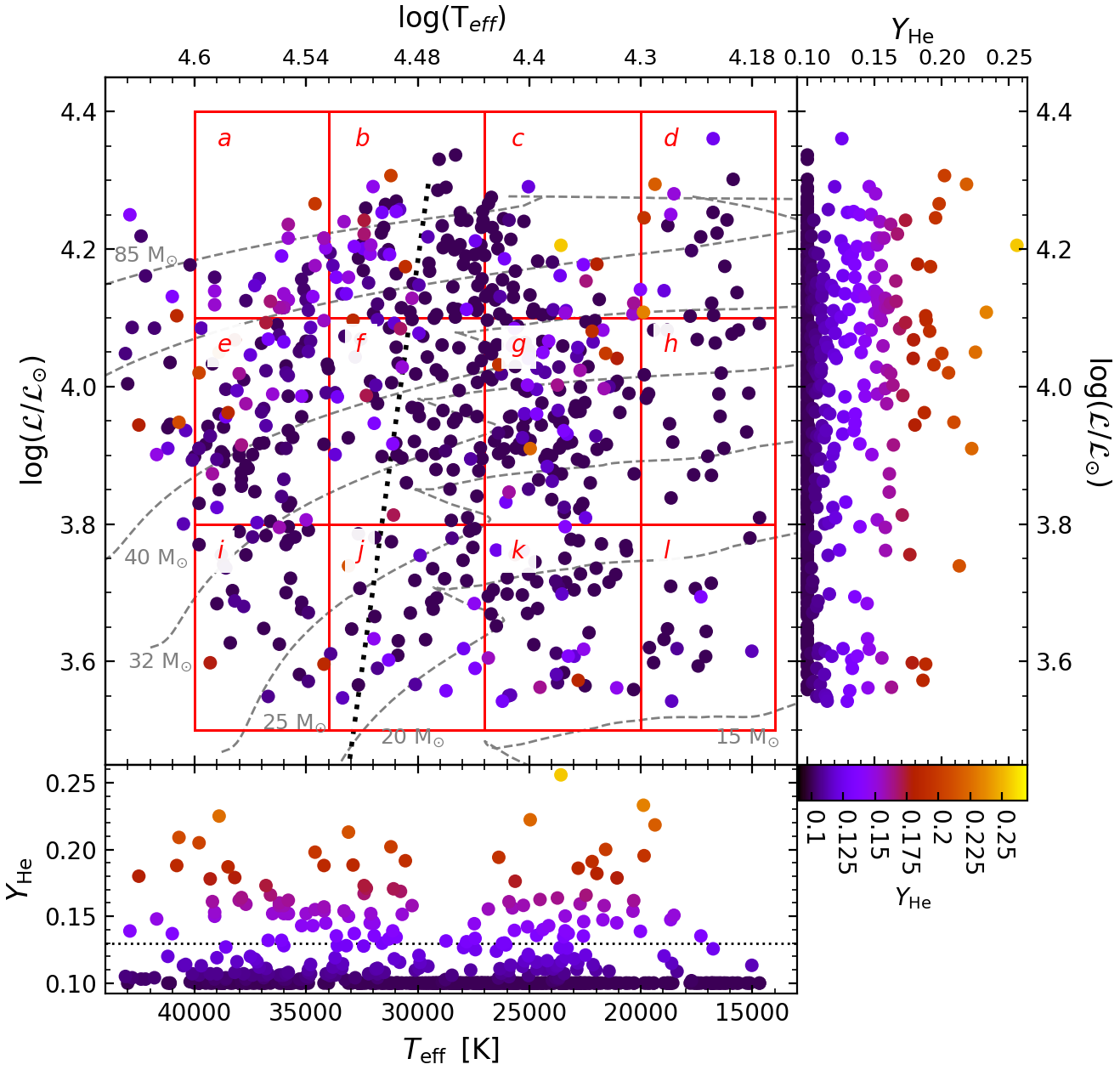}}
\caption{sHR diagram of the stars in the sample plus 191 O-type stars from Hol18-22, color-coded by helium abundance. We adopted \He$\approx$\,0.10 as the lowest possible value. The bottom and right sub-panels show \He against \Teff and log\,\Ls, respectively. The various sub-groups of stars listed in Table~\ref{tab:he_percentages} are indicated with red solid lines. Results considered as lower limits or degenerate are excluded. The bottom sub-panel includes a horizontal dotted line at \He=\,0.13.} Evolutionary tracks are the same as in Fig.~\ref{fig:shrd}. 
\label{fig:shrd_he}
\end{figure}

Figure~\ref{fig:shrd_he} shows the sHR diagram for the estimated surface abundances of helium in our BSG and O-star sample. As in previous similar figures, we also present two sub-panels to investigate potential dependencies between this quantity and log\,\Ls and \Teff, respectively.

Globally speaking, our results cover the range \He=\,0.10\,--\,0.23, with few exceptions. For the discussion below, and based on the median of our results plus the average of all the error estimates for our sample stars, we define \He=\,0.13 as the threshold for a star to be considered He-enriched. We find that 20\% of the stars in our sample have a surface helium abundance above this limit, of which only 7\% display \He>\,0.16. 

The right and bottom sub-panels of Fig.~\ref{fig:shrd_he} do not show any clear correlation between the amount of He surface enrichment and log\,\Ls or \Teff. To further investigate the potential correlation between these three quantities, also taking into account the \vsini of the stars, Table~\ref{tab:he_percentages} summarizes some information of interest regarding the percentages of stars with \He>\,0.13. This information is associated with subsamples of stars located within the 12 panels highlighted in red in Fig.~\ref{fig:shrd_he}. Specifically, we have selected three ranges in \Teff that presumably cover the MS (see Sect.~\ref{subsection:51_tmp}), plus a fourth one corresponding to stars with \Teff<\,20\,kK (i.e., to the cooler side of the suggested empirical TAMS).

Regarding stars with \Teff>\,20\,kK, the most remarkable result is the particularly large percentage of He-enriched stars in panel $a$ (reaching $\approx$60\%); all other panels with \Teff>\,20\,kK show only $\approx$10 to 20\% of He-enriched objects, again without any correlation between this quantity and \Teff or log\,\Ls. The statistics associated with the rightmost panels in Fig.~\ref{fig:shrd_he} ($d$, $h$, and $l$) show a different behavior, with a much lower percentage of He-enriched stars (except for panel $d$).

\begin{table}[!t]
\centering
\caption[]{Summary properties of He-enriched stars, separated in groups of different \Teff and \logLs, as indicated in the second and third columns. The first column refers to the panels displayed in Fig.~\ref{fig:shrd_he}.} 
 \label{tab:he_percentages}
 \begin{tabular}{cccc|ccc}
   \hline
   \hline
   \noalign{\smallskip}
 \multirow{3}{*}{\rotatebox[origin=c]{90}{Panel}} 
 &$\log (\sfrac{\mathcal{L}}{\mathcal{L}_{\odot}})$
 &\Teff &All   &\multicolumn{3}{c}{He-enriched (\He>\,0.13)\,$^{a}$} \\
 &range &range &\#  &All &low\,-- &high\,-- \\
 &[dex] &[kK]  &    &    &\vsini  &\vsini   \\
   \noalign{\smallskip}
   \hline
   \noalign{\smallskip}
   $a$ & \multirow{4}{*}{\rotatebox[origin=c]{90}{4.35\,--\,4.10}} 
         & 40\,--\,34 & 25 & 62\%& 57\% & 64\% \\
   $b$ & & 34\,--\,27 & 78 & 19\%& 16\% & 25\% \\
   $c$ & & 27\,--\,20 & 42 & 21\%& 24\% &  0\% \\
   $d$ & & 20\,--\,14 & 18 & 26\%& 26\% &  -   \\
   \noalign{\smallskip}
 \hline
   \noalign{\smallskip}
   $e$ &  \multirow{4}{*}{\rotatebox[origin=c]{90}{4.10\,--\,3.85}} 
         & 40\,--\,34 & 69 & 18\%& 12\% & 22\% \\
   $f$ & & 34\,--\,27 & 63 & 10\%&  0\% & 26\% \\
   $g$ & & 27\,--\,20 &106 & 15\%& 12\% & 21\% \\
   $h$ & & 20\,--\,14 & 21 &  0\%&  0\% &  -   \\
   \noalign{\smallskip}
 \hline
   \noalign{\smallskip}
   $i$ &  \multirow{4}{*}{\rotatebox[origin=c]{90}{3.85\,--\,3.60}} 
         & 40\,--\,34 & 36 & 10\%&  5\% & 20\% \\
   $j$ & & 34\,--\,27 & 34 & 19\%& 14\% & 29\% \\
   $k$ & & 27\,--\,20 & 47 & 18\%& 11\% & 38\% \\
   $l$ & & 20\,--\,14 & 14 &  7\%&  8\% &  0\% \\
   \noalign{\smallskip}
 \hline
 \end{tabular}
 \begin{list}{}{}
  \item {\bf Notes}. $^{(a)}$ Low- and high-\vsini refer to stars with \vsini below or above 100\kms, respectively. The indicated percentages have been computed with respect to the total number of stars in each \vsini subgroup within the corresponding panel. $^{(b)}$ Abundances from \citet{holgado19PhD} are used for the 191 O-stars shown in Fig.~\ref{fig:shrd_he}.
 \end{list}
 \end{table}

Another interesting result is that in those panels where there is a clearly bimodal \vsini distribution (namely those with \Teff>\,20\,kK, see bottom sub-panel of Fig.~\ref{fig:shrd_he}), the percentage of He-enriched stars in the tail of fast rotators is systematically higher (except for panel $c$) than for the main low \vsini component. Indeed, a two-sample Kolmogorov-Smirnov test indicates that, with a 95\% confidence, both groups (now also considering the non-He-enriched stars) do not arise from the same probability distribution for the surface helium abundance. This might be explained by attributing a different origin to the He-enriched stars in both low-\vsini and fast-rotating stellar populations.

While we expect these results to serve as guidelines for future, in-depth comparisons of single and binary evolution model predictions, we provide here some first hints which can be extracted from the information in Table~\ref{tab:he_percentages}. First, we evaluate the possibility that He-enriched stars originate from single-star evolution. For this, we compare with evolutionary model predictions from \citet{brott11}, \citet{ekstrom12}, and \citet{keszthelyi22}. Among them, only the models by \citet{ekstrom12} with an initial rotational velocity of 40\% of critical rotation can explain some (but certainly not all, see below) of the percentages of stars with \He>\,0.13 quoted in Table~\ref{tab:he_percentages}. The alternative computations by \citet{brott11}, and \citet{keszthelyi22} do not produce any remarkable He-enrichment along those main sequence tracks crossing any of the various red panels highlighted in Fig.~\ref{fig:shrd_he}.

Exploring further the evolutionary models computations by \citet{ekstrom12}, we have found that they can, at maximum, explain 40\,--\,50\% of the detected stars with He-enriched surfaces. Basically, these are targets with low and intermediate projected rotational velocities (\vsini<\,100\,--\,150\,\kms) in panels $a$, $b$, $e$, and $f$. Whereas the observed percentage of stars with He-enriched surfaces is systematically larger within the tail of fast rotators (see above and Table~\ref{tab:he_percentages}), \citet{ekstrom12}, on the other hand, predict that those stars with a clearly detected enrichment of helium should have also suffered from a significant braking of the stellar surface.

\begin{figure}[!t]
\centering
\resizebox{\columnwidth}{!}{\includegraphics{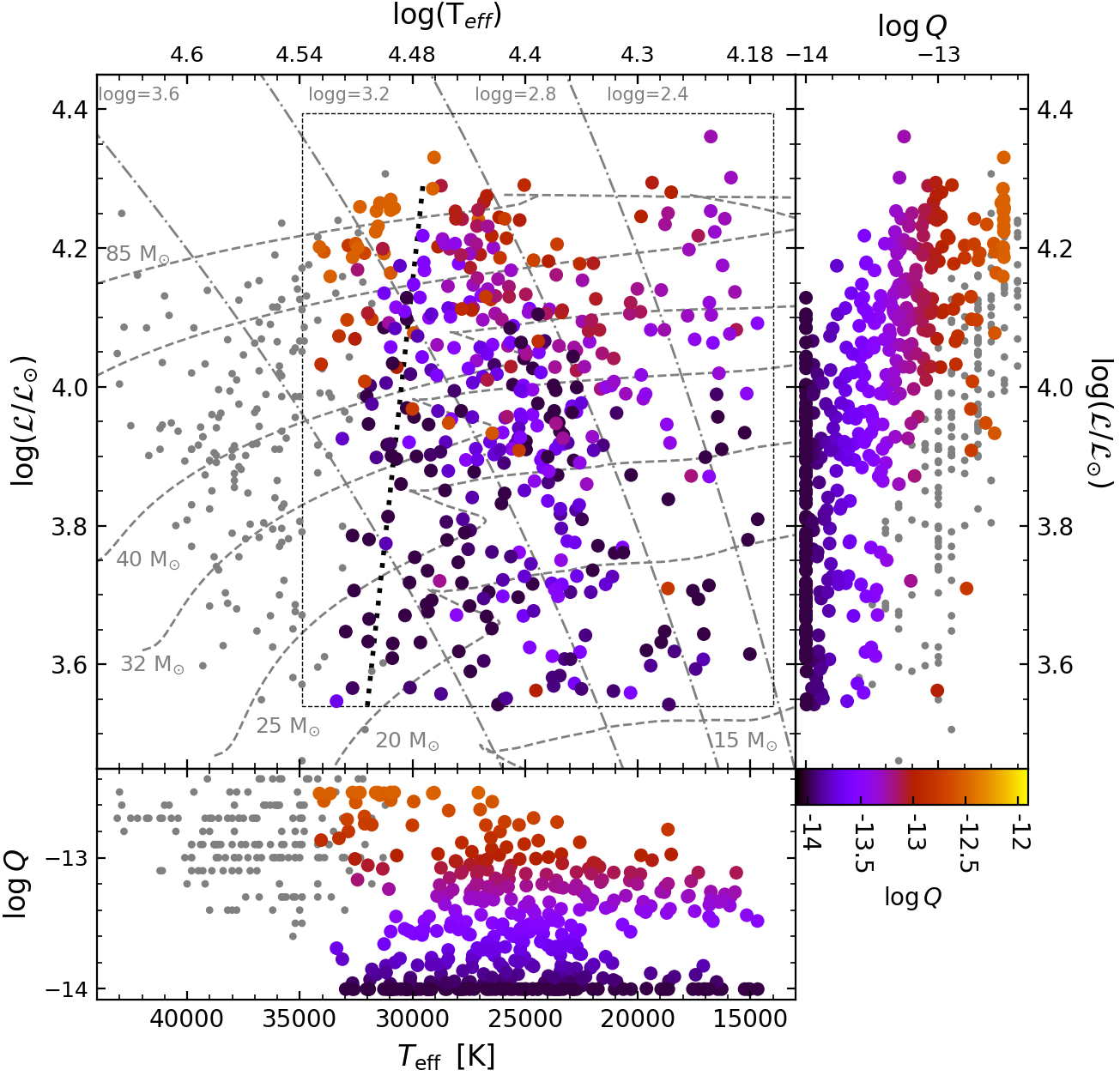}}
\caption{sHR diagram of the stars in the sample color-coded by the wind-strength parameter. The bottom and right sub-panels in each panel show this quantity against \Teff and log\,\Ls, respectively. Cases in which \logQ is degenerate are excluded (see Sect.~\ref{subsection:41_tmp}). All panels include 191 O-type stars from Hol18-22 indicated with gray circles. Evolutionary tracks are the same as in Fig.~\ref{fig:shrd}.} 
\label{fig:shrd_logq}
\end{figure}

All this, together with the increasing empirical evidence indicating that main-sequence massive stars might not be suffering from such a significant surface braking (see Sect.~\ref{subsection:52_tmp}) leaves us with the necessity for an alternative scenario to explain an important fraction (if not all) of the detected He-enriched stars in our sample, particularly those with \vsini>\,150\,\kms.

In this context, given the high percentage of massive stars born in binary and multiple systems, and the high probability of an interaction during their evolution \citep[see][and references therein]{marchant23}, stars that exhibit helium surface enrichment might be the result of binary interaction. For fast-rotating objects, they could be the gainers of post-interaction systems in which the mass transfer event occurs when the initially more massive star has evolved beyond the MS \citep[i.e. case B mass transfer][]{langer20, wang20, klencki20, sen22}. Moreover, some of the He-enriched low-\vsini stars could be the products of merger events, including cases in which the merging occurs when one or both components are close to or beyond the TAMS \citep[see][]{podsiadlowski92, langer12, schneider16}. Therefore, a more thorough investigation of the various possibilities opened by the binary channel, incorporating information about C, N, and O surface abundances and new predictions from single and binary evolutionary models, is hence certainly needed.


\subsection{Wind properties}
\label{subsection:55_tmp}

\begin{figure}[!t]
\centering
\resizebox{\columnwidth}{!}{\includegraphics{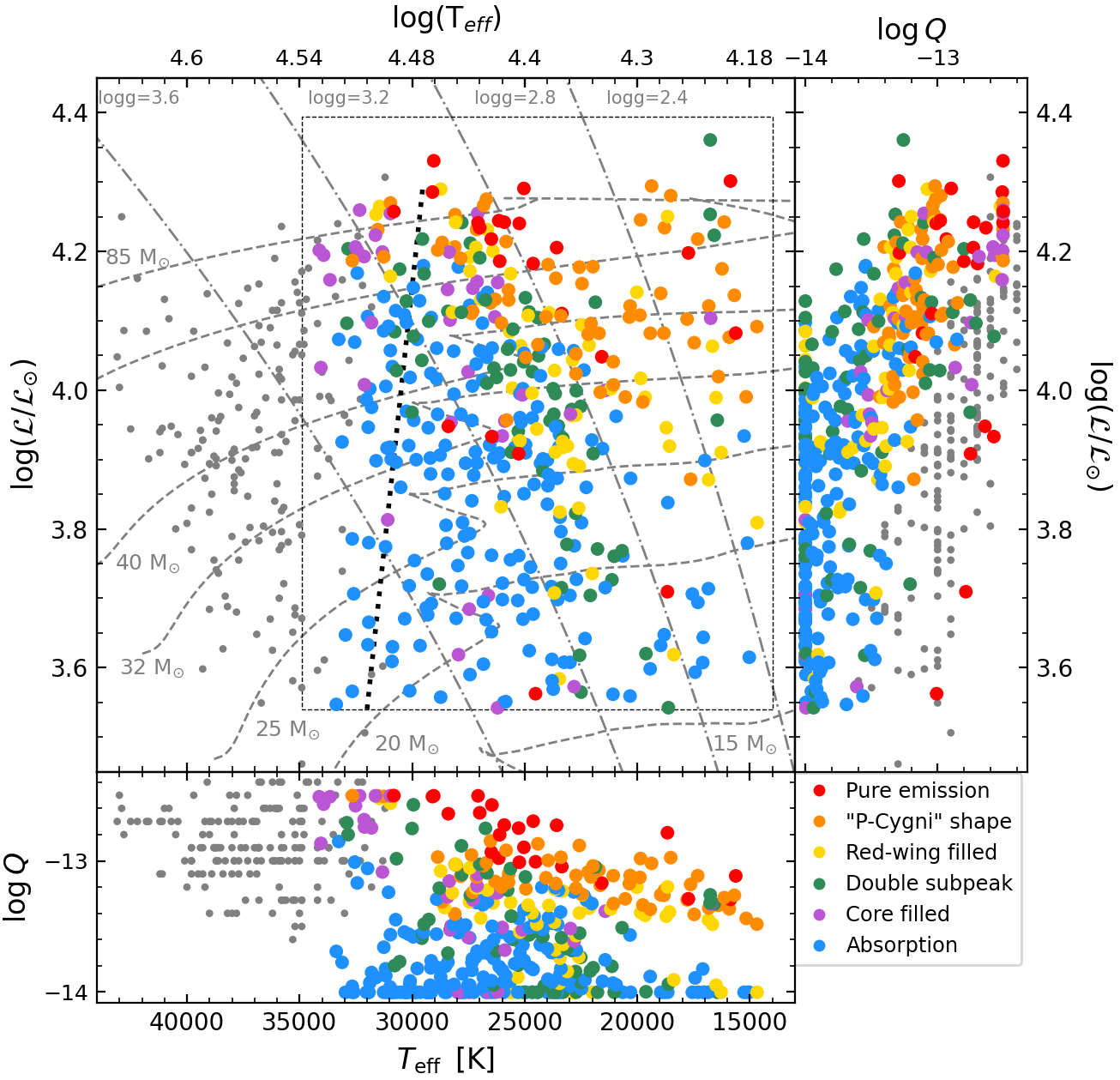}}
\caption{Same as Fig.~\ref{fig:shrd_logq}, but color-code representing the different shapes of the H$\alpha$ line as classified in \citet[][see also labels within the bottom right inset]{deburgos23}.} 
\label{fig:shrd_logq_sep}
\end{figure}

At present, several NLTE atmospheric codes are able to treat spherically extended atmospheres with winds (assuming radiative equilibrium). In this work we used {\tt FASTWIND} (see Sect.~\ref{subsubsection:321_tmp}), but other available codes are {\tt CMFGEN} \citep{hillier98}, {\tt PoWR} \citep{grafener02, hamann04}, {\tt WM-basic} \citep{pauldrach01}, or {\tt PHOENIX} \citep{hauschildt92}. A major challenge in reproducing the observed spectral lines affected by stellar winds is accounting for the inhomogeneities (clumping) of these winds \citep[see][and references therein]{puls08}. Such inhomogeneities can only be described by adopting a large number of free parameters \citep[particularly, when modeling optically thick clumping, see,][]{sundqvist18}, which significantly increase the complexity. However, recent studies have gradually tried to improve this scenario \citep[see, e.g.,][]{hawcroft21, brands22, pernini-peron23}, since empirical constraints are key to derive important wind properties such as mass-loss rates.

In this work, we limit the discussion of such wind properties to our results for the wind-strength parameter\footnote{here derived adopting an unclumped wind; however, when replacing $\dot{M}$ by $\dot{M} \sqrt{f_{\rm cl}}$  in the definition of $Q$ (with $f_{\rm cl}$ the conventional clumping factor for optically thin clumping), the $Q$-values derived in this work remain roughly valid also for inhomogeneous winds.} and its relation to the morphology of the H$\alpha$ line. Further, more detailed investigations on the actual mass-loss rates and clumping properties will be presented in a forthcoming study.

Figure~\ref{fig:shrd_logq} displays the stars from Fig.~\ref{fig:shrd}, now colored by the wind-strength parameter. The results from Hol18-22 have been included for reference (in gray). The bottom sub-panel shows two main and important features. First, our results do not show evidence for increasing mass-loss rates over the bi-stability region towards lower \Teff. Instead, we observe a slow decay of the maximum \logQ values, with \logQ$\approx$\,$-13$ in the 20\,--\,25\,kK range. Second, we find a clear separation of two groups of stars below $\approx$22\,kK, one with \logQ\gs$-13.6$, and another one at (or below) \logQ$\approx$\,$-14.0$. We will return to this bimodal distribution later, when discussing the observed H$\alpha$ morphology. Moreover, a diagonal gap dividing O- and B-type stars seems to be present.

The right-hand sub-panel displays increasing \logQ values with increasing log\,\Ls. This feature is expected since stars closer to the Eddington limit should and indeed do possess stronger stellar winds driven by intense radiation \citep[see, e.g.,][]{abbott80, pauldrach86}. We also note that the wind-strengths of stars with \logLs\ls3.9\,dex are considerably weaker (\logQ\ls$-13.25$) than those for values above.

Due to our neglect of wind inhomogeneities, discrepancies between the synthetic spectra from our best-fitting models and observations are to be expected, at least if the clumping properties would vary as a function of location \citep[which seems to be the case, e.g.,][]{najarro11}. In fact, these neglected inhomogeneities are the most likely contributors to the larger $\chi^{2}$ values associated with the H$\alpha$ line (see Fig.~\ref{fig:chi2_quality}). Interestingly, the majority of cases where H$\alpha$ could not be reproduced by our modeling correspond to profiles either displaying emission in both line wings or a P-Cygni shape with very strong emission in the red wing. In some of these cases, the models were also unable to reproduce the shape of H$\beta$ if it was not in pure absorption.

These results significantly increase the number of luminous blue stars for which wind-densities have been derived, compared to previous studies \citep[see, e.g.,][]{markova08, haucke18}.

To enable an investigation of the relation between wind-strength parameter and line-profile morphology of typical wind lines, in \citet{deburgos23} we carried out a visual classification of the shape of the H$\alpha$ and H$\beta$ line profiles. We accounted for six different line profiles: ``Pure emission" profiles when the profile is in emission above the normalized flux, ``P-Cygni shape" profiles when the emission is only in the red part of the line profile, ``red filling" profiles when the red wing of the line is filled up to the continuum, ``double subpeak" profiles when both wings of the line are filled or in emission above the normalized flux, ``core filled" profiles when the core is filled to some degree, and ``absorption" profiles when the line is in absorption. Using this classification, Fig.~\ref{fig:shrd_logq_sep} shows, for the first time, the morphological map for H$\alpha$ in the sHR diagram for our BSG sample. The central panel shows a gradient of profile types towards lower \Teff and higher log\,\Ls, from absorption profiles to profiles with double subpeak, to profiles where the red-wing is filled or in emission, to those cases with pure emission. This gradient agrees very well with our previous findings in \citet{deburgos23} using the spectral classifications. 

Another, even more interesting feature is displayed in the bottom sub-panel: here, the separation of stars above and below \logQ$\approx-13.6$ (cf. Fig.~\ref{fig:shrd_logq}) is even more evident, since stars from each group differ significantly regarding their H$\alpha$ morphology. In particular, the low-\logQ group consists of absorption profiles, whereas those with large \logQ mostly comprise ``P-Cygni shape" profiles. This separation is also present with respect to LC. Those stars above \logQ$\approx-13.6$ all correspond to Ia luminosities, whereas for those other stars below \logQ$\approx-13.6$, the majority corresponds to Ib and II.

On the other hand, from the central panel, we see that the majority of core-filled profiles are located at spectral type O9 and \logLs$\approx$\,4.2 dex (above the absorption profiles). Similarly, most of the profiles exhibiting pure emission are concentrated around B0\,I type stars. Moreover, we also note the presence of few stars with H$\alpha$ in pure emission, located at \logLs<\,3.8\,dex, where absorption profiles dominate. Their location in the sHR diagram suggests that these objects might be Be stars.


\section{Concluding remarks}
\label{section:6_tmp}

We have conducted a quantitative spectroscopic analysis of high-resolution and high signal-to-noise optical spectra of 527 Galactic O9\,--\,B5 stars, collected from the IACOB spectroscopic database and the ESO public archive. The outcome of our analysis represents the most extensive collection of homogeneously determined spectroscopic parameters of Galactic BSG stars built to date, superseding previous attempts by more than one order of magnitude. This study aims to advance our understanding of the evolutionary nature of massive stars, also establishing new empirical anchor points for state-of-the-art and future model computations.

The spectroscopic analysis was carried out in two steps. First, we used {\tt IACOB-BROAD} to derive the projected rotational velocity and macroturbulent broadening. Second, we used a suitable grid of model atmospheres computed with the {\tt FASTWIND} code to create a statistical emulator for {\tt FASTWIND} synthetic spectra. In combination with a Markov chain Monte Carlo method, we derived the fundamental atmospheric parameters, helium and silicon surface abundances, as well as an indicator of the wind strength.  

We present a revised calibration of \Teff against spectral type for Galactic B-type supergiants (LC\,I) down to B5-type stars. Previous calibrations based on smaller samples differ by up to 3\,kK for some SpT bins when comparing a third-order polynomial fit to the data. Reliable spectral classifications from selected sources turned out to be crucial to avoid spurious results. In this latter regard, SIMBAD classifications for B-type stars exhibit inaccuracies, emphasizing the need for a systematic and reliable revision, akin to efforts in O-type star studies. 

In comparison with the O-type stars, the relatively large number of early B-type supergiant stars included in our magnitude-limited sample suggest that at least a non-negligible fraction of them could still be on the MS, in contrast to the classic interpretation of these being He-core burning post-MS objects. Our results present solid statistical evidence for a drastic drop in the relative number of objects at \Teff$\approx$\,21\,kK. Though further analyses are certainly required, we suggest that this drop (roughly occurring at B2-type stars) empirically locates the TAMS in the mass range between 15 and 85\MSol.

Similarly to O-type stars, the distribution of projected rotational velocities for evolved B-type stars also exhibits two clear components. Namely, a low \vsini (\ls100\kms) component that is present in the full range of covered effective temperatures, and a tail of fast rotators (reaching \vsini values up to $\approx$400\kms) which disappears below $\approx$21\,kK. Guided by some recent theoretical scenarios, our empirical study is consistent with the possibility that this tail of fast rotators is mostly populated by post-interaction binary products.

We observe no surface braking in the low \vsini component along the whole considered range of effective temperatures. This result, combined with a constant percentage of stars with \vsini>\,100\kms and associated maximum \vsini values from the ZAMS to the drop, might indicate the existence of a very efficient angular momentum transport mechanism between the core and the surface of massive stars.

Whereas the scarcity of stars populating the tail of fast rotators below \Teff$\approx$\,21\,kK has been attributed to either the end of the MS or the result of an enhanced angular momentum loss at the theoretically predicted bi-stability jump, our combined results disfavor the latter scenario. 

The distribution of microturbulent velocities in the BSG domain is for the first time presented in an sHR diagram. A strong correlation between \vmic and log\,\Ls is found. In agreement with previous findings, a decrease of both \vmic and \vmac towards lower \Teff is also observed. This might indicate a connection between the physical mechanisms responsible for both {\em turbulent} motions; our sample of stars supports higher \vmic values to be associated with higher \vmac. 

Our findings for the helium surface abundance indicate that, on average, only $\approx$20\% of luminous blue stars show helium enrichment (\He>\,0.13) in their atmospheres. No clear correlation is found between the surface abundance of helium and log\,\Ls or \Teff. However, while we find a significantly lower percentage of He-enriched stars on the cooler side of the suggested empirical TAMS (\Teff\ls\,20\,kK), the percentage of He-enriched stars in the tail of fast rotators is systematically higher than for the main low \vsini component for stars with \Teff above $\approx$20\,kK. In addition, we show with high statistical confidence that both groups of different \vsini do not originate from the same probability distribution for the surface helium abundance, suggesting a different physical origin of both populations.

Compared with predictions from state-of-the-art evolutionary models, and considering the empirical evidence that the predicted surface braking might not occur, the possibility that He-enriched stars originate from single-star evolution seems less likely compared to a binary evolution origin.

Last, we evaluated the wind-strength parameter and its correlation with the morphology of the H$\alpha$ profiles. Our results indicate no evidence of a mass-loss increase over the expected wind bi-stability region, but rather a slow decay of the maximum \logQ values with decreasing \Teff. We found a separation of stars with \logQ above and below $-13.6$ in the \Teff range below $\approx$22\,kK, each one displaying a different H$\alpha$ morphology. 
The presence of a positive correlation between \logQ and log\,\Ls is also evident, where the highest values are concentrated at \logLs\gs3.9\,dex. 
In general, we found a gradient of morphologies for H$\alpha$ across the sHR diagram, with some profile shapes being concentrated in specific areas in the diagram.

As a final remark, the results presented here represent a significant step forward in the empirical spectroscopic study of Galactic luminous blue stars, providing an updated overview of many of their properties. These findings also lay the groundwork for forthcoming in-depth studies dedicated to specific properties of our sample. For this, additional information is certainly required on luminosities, masses, radii, surface elemental abundances, and wind properties. Our ultimate objective is to establish new empirical anchor points that can serve to improve our understanding of the evolutionary nature of BSGs.


\begin{acknowledgements}

AdB and SS-D acknowledge support from the Spanish Ministry of Science and Innovation (MICINN) through the Spanish State Research Agency through grants PID2021-122397NB-C21, and the Severo Ochoa Programme 2020-2023 (CEX2019-000920-S).
The authors would like to thank Z. Keszthelyi, D. Lennon, and N. Przybilla for their useful and valuable comments, and I. Negueruela for providing us with revised spectral classifications.
We give special thanks to all the observers who contributed to the acquisition of the spectra used here in this work. Among them, especially to G. Holgado and J. Maíz-Apellániz.
Regarding the observing facilities, this research is based on observations made with the Mercator Telescope, operated by the Flemish Community at the Observatorio del Roque de los Muchachos (La Palma, Spain), of the Instituto de Astrof\'isica de Canarias. In particular, obtained with the HERMES spectrograph, which is supported by the Research Foundation - Flanders (FWO), Belgium, the Research Council of KU Leuven, Belgium, the Fonds National de la Recherche Scientifique (F.R.S.-FNRS), Belgium, the Royal Observatory of Belgium, the Observatoire de Genève, Switzerland and the Thüringer Landessternwarte Tautenburg, Germany.
This research also based on observations with the Nordic Optical Telescope, owned in collaboration by the University of Turku and Aarhus University, and operated jointly by Aarhus University, the University of Turku and the University of Oslo, representing Denmark, Finland and Norway, the University of Iceland and Stockholm University, at the Observatorio del Roque de los Muchachos, of the Instituto de Astrof\'isica de Canarias.
Additionally, this work has made use of observations collected from the ESO Science Archive Facility under ESO programs: 60.A-9700(A), 72.D-0235(B), 73.C-0337(A), 73.D-0234(A), 73.D-0609(A), 74.D-0008(B), 74.D-0300(A), 75.D-0103(A), 75.D-0369(A), 76.C-0431(A), 77.D-0025(A), 77.D-0635(A), 79.A-9008(A), 79.B-0856(A), 81.A-9005(A), 81.A-9006(A), 81.C-2003(A), 81.D-2008(A), 81.D-2008(B), 82.D-0933(A), 83.D-0589(A), 83.D-0589(B), 85.D-0262(A), 86.D-0997(B), 87.D-0946(A), 88.A-9003(A), 89.D-0975(A), 90.D-0358(A), 91.C-0713(A), 91.D-0061(A), 91.D-0221(A), 92.A-9020(A), 95.A-9029(D), 97.A-9039(C), 102.A-9010(A) and 179.C-0197(C).

\end{acknowledgements}

\typeout{}
\bibliographystyle{aa} 
\bibliography{biblio} 

\begin{appendix}


\section{Grid of input models}
\label{apen.input_maui}

\begin{figure}[!h]
\centering
\resizebox{\columnwidth}{!}{\includegraphics{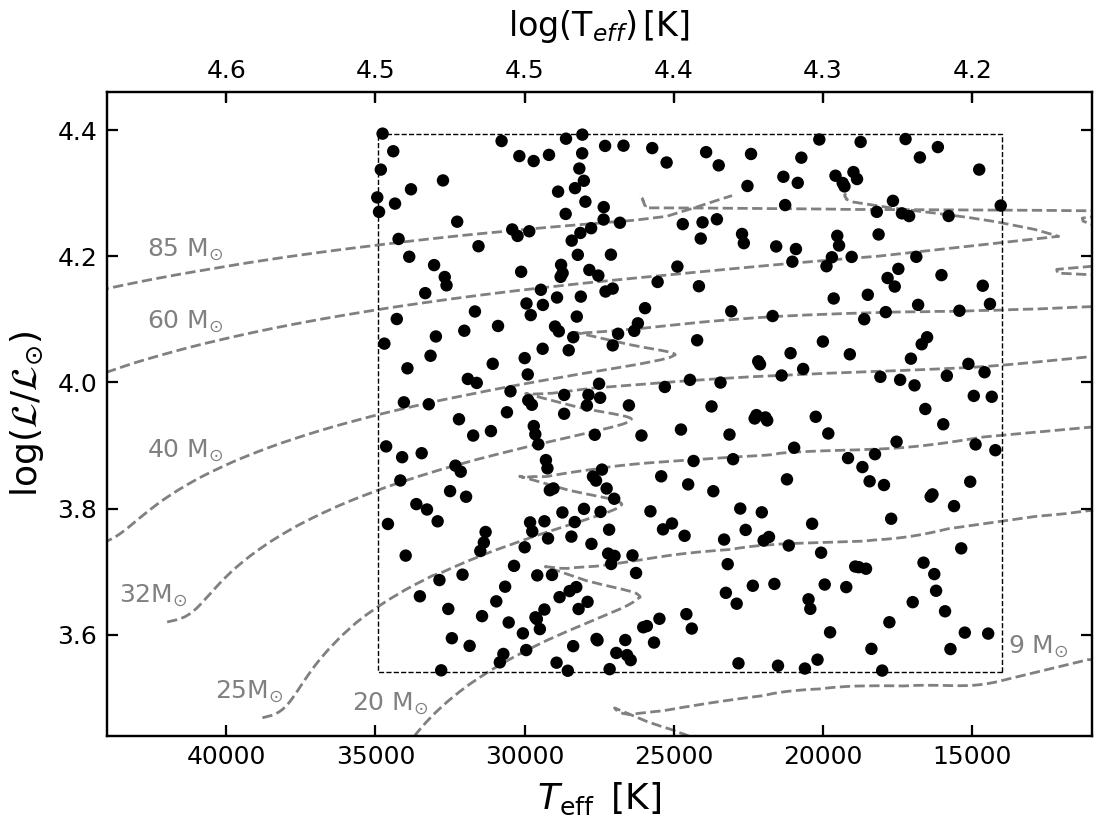}}
\caption{sHR diagram showing the initial grid of 358 {\tt FASTWIND} computed models used in this work. The boundary limits are marked with dashed black lines. A set of Geneva non-rotating evolutionary tracks with solar metallicity is included for reference.}
\label{fig:train_models}
\end{figure}

\begin{figure}[!t]
\centering
\resizebox{\columnwidth}{!}{\includegraphics{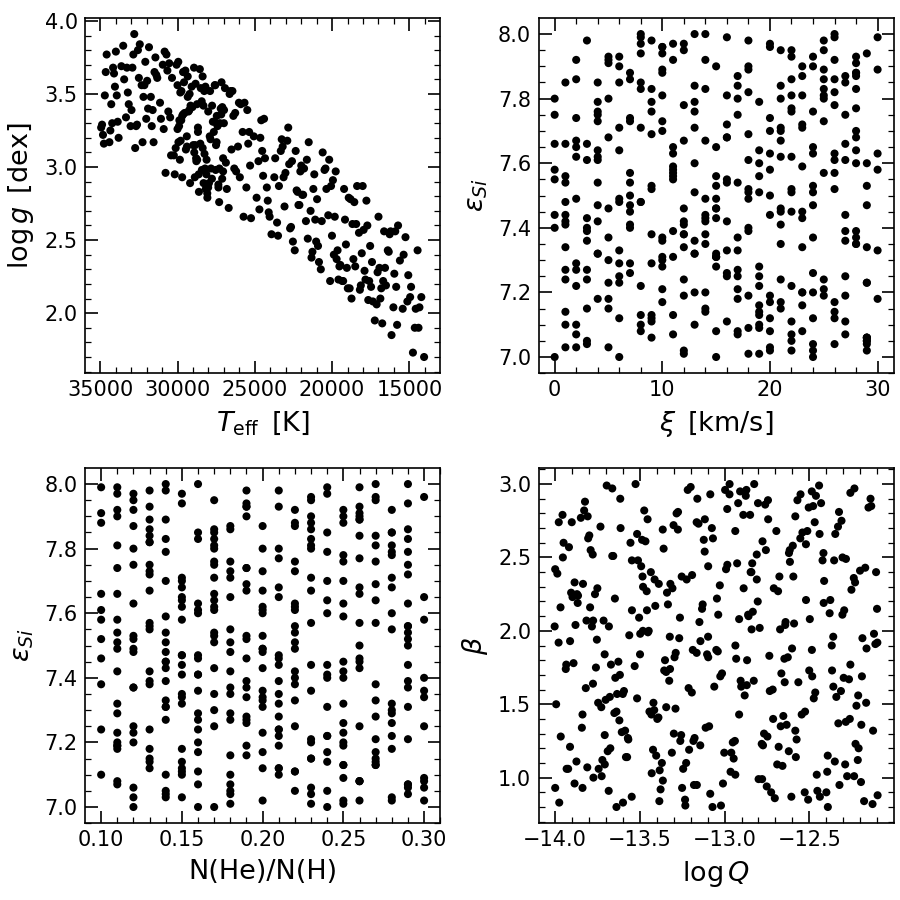}}
\caption{Same {\tt FASTWIND} models as in Fig.~\ref{fig:train_models} but now displaying the coverage of other stellar and wind parameters.}
\label{fig:train_models_others}
\end{figure}

Figure~\ref{fig:train_models} displays an sHR diagram with the grid of 358 {\tt FASTWIND} models used to train the statistical emulator which, as described in Sect.~\ref{subsubsection:322_tmp}, is used to reproduce equivalent {\tt FASTWIND} simulations. Figure~\ref{fig:train_models_others} illustrates the coverage of these models with respect to the remaining spectroscopic parameters. 

\section{Visualization of the output solution}
\label{apen.output_maui}

\begin{figure*}[!t]
\centering
\includegraphics[width=1\textwidth]{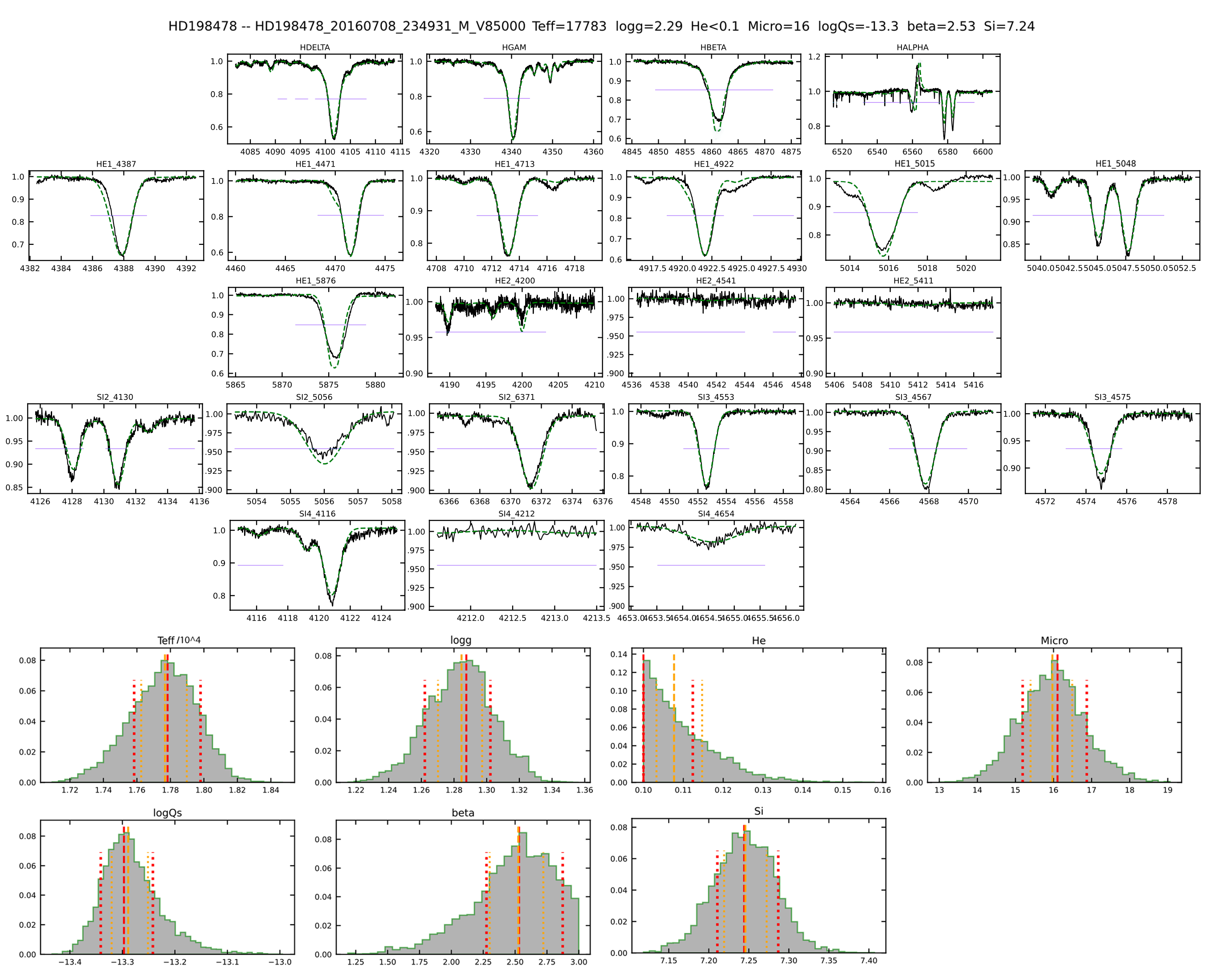}
\caption{Summary of the analysis for HD\,198\,478. The sub-panels of the top five rows show the observed spectra (solid black line) and synthetic model (dashed green line) in different diagnostic windows used in the analysis of our sample. Within each window, the purple horizontal line indicates which sub-regions have not been masked in the analysis. The sub-panels in the bottom two rows are the associated probability distributions of each of the parameters derived in this work. The vertical dashed red and orange lines indicate the maximum and median of the distribution and associated uncertainties.}
\label{fig:output_maui}
\end{figure*}

\begin{figure*}[!t]
\centering
\includegraphics[width=1\textwidth]{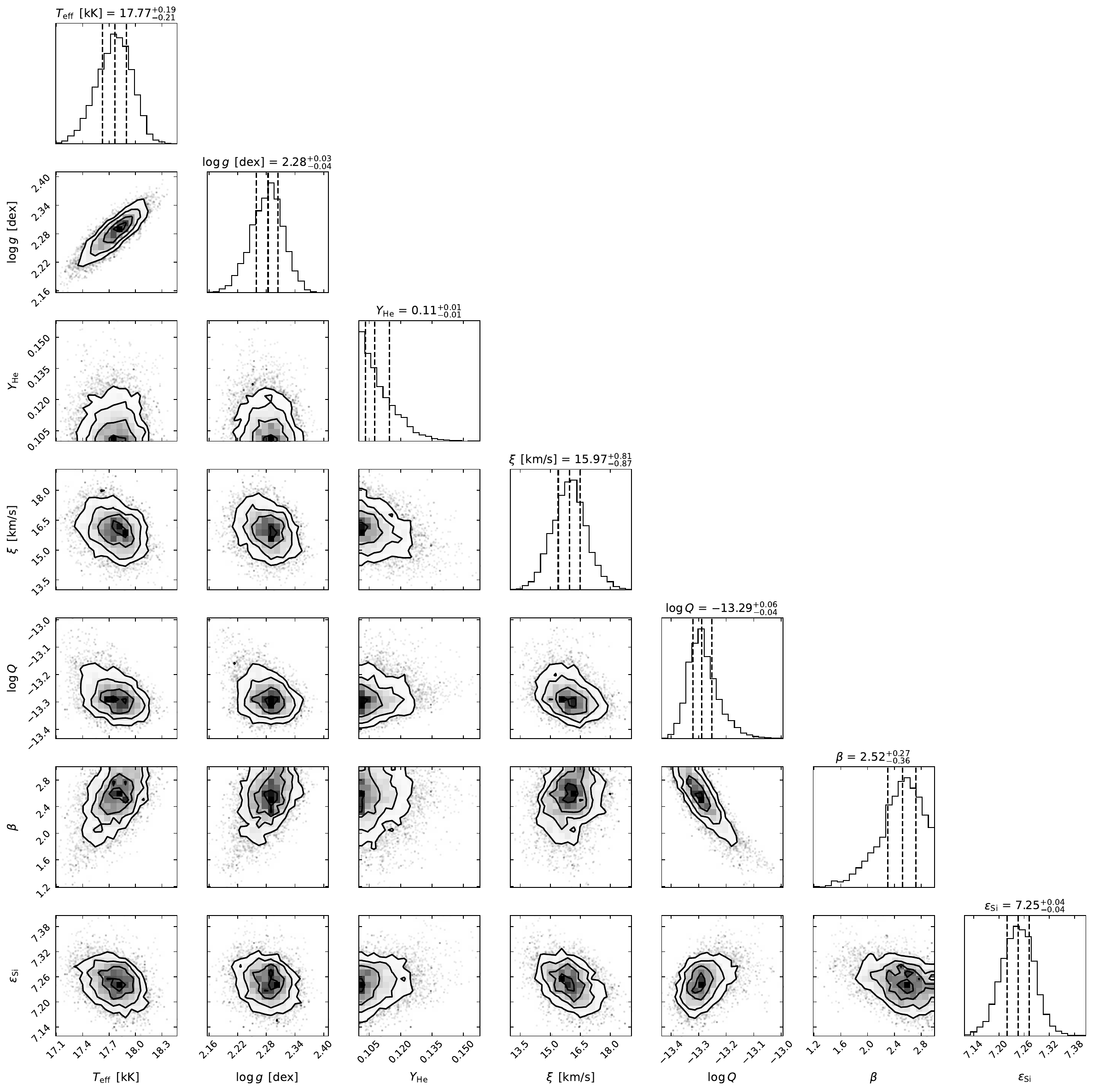}
\caption{Corner plot with the PDFs of the seven parameters derived for HD\,198\,478. }
\label{fig:corner_plot}
\end{figure*}

In Sect.~\ref{subsection:32_tmp}, we described the method used to derive the best (i.e., most probable) set of parameters for each star. In each analysis, an associated synthetic spectrum is obtained, together with individual probability distributions of the parameters. An example of this output is included in Fig.~\ref{fig:output_maui} for HD\,198\,478. The top array of sub-panels shows the observed and synthetic spectra split into different windows that include the diagnostic lines listed in Table~\ref{tab:diag_lines}. The associated probability distributions are shown in the bottom grid of sub-panels. Complementary to that figure, Fig.~\ref{fig:corner_plot} shows, for the same star, the probability distributions of the different parameters in a ``corner plot", allowing to visualize the possible covariances. In this example, we see the (well-known) presence of a significant covariance between \Teff and \logg, arising from the behavior of the H-lines, as well as between $\beta$ and \logQ \citep[see][]{markova04}. In addition, we can see that the distribution of the helium surface abundance reaches the lower boundary of the grid.

\begin{figure*}[!t]
\centering
\includegraphics[width=1\textwidth]{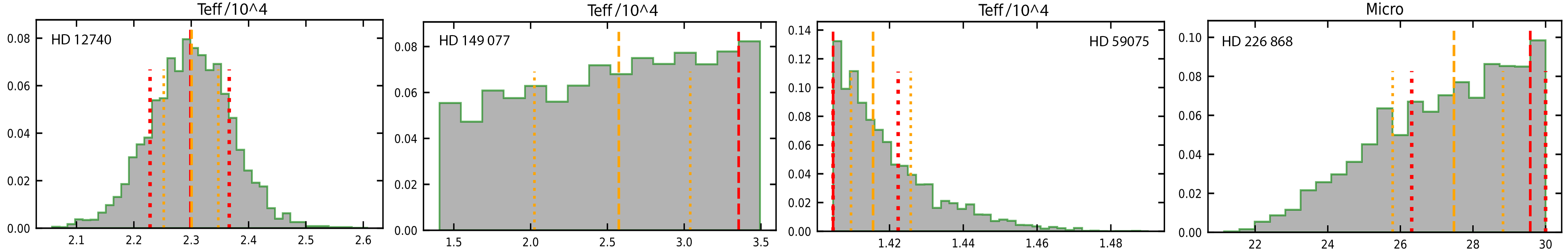}
\caption{Illustrative example of the four different PDFs described in Sect.~\ref{subsubsection:324_tmp}. From left to right: a well-behaved distribution (case $a$); an example of an undefined (degenerate) solution ($d$); and two cases for the upper and lower limits ($c$, $b$), respectively.}
\label{fig:4_distributions}
\end{figure*}

For each of the stellar and wind parameters, we also defined in Sect.~\ref{subsubsection:324_tmp} four possible scenarios for the associated probability distribution. Figure~\ref{fig:4_distributions} provides an example for each of the four possible cases described there.

\section{New spectral classifications}
\label{apen.dwarfs}

Table~\ref{tab:newclass_dwarfs} includes revisited spectral classifications based on visual morphological features for a group of stars erroneously classified as dwarfs in the default SIMBAD classification.

\begin{table}[ht]
\caption{New spectral classifications for some stars erroneously classified as dwarfs, taken from Fig.~\ref{fig:shrd}}.
\label{tab:newclass_dwarfs}
    \centering
    \begin{tabular}{lll}
        \hline\hline\noalign{\medskip}
        ID & SpC$_{\,SIMBAD}$ & SpC$_{\,This\,work}$ \\
        \noalign{\vspace{0.2cm}}\hline\noalign{\smallskip}
        \multicolumn{3}{c}{Stars erroneously classified as dwarfs}\\
        \noalign{\smallskip}\hline\noalign{\smallskip\smallskip}\smallskip
        HD~13544   & B0.5~IV  & B1~IIn     \\\smallskip 
        HD~180968  & B3~IV+   & B1~IIIn    \\\smallskip 
        HD~193444  & B0.5~V   & B0.5~III   \\\smallskip 
        HD~201819  & B1~Vp    & B1~IIIn    \\\smallskip 
        HD~35653   & B0.5~V   & B1~II      \\\smallskip 
        HD~43703   & B1~IVp(e)& B1~IIIn    \\ 
        \hline
    \end{tabular}
    \tablefoot{The default SIMBAD classification is included for comparison.}
\end{table}


\section{Long tables}

\onecolumn
\scriptsize{
\begin{landscape}
\begin{longtable}{lcccccccrcrcrcclc}
\label{tab:qsa_results}\\
\caption[]{Results of the spectroscopic analysis of the stars in the sample (extract)}\\

 \hline\hline
 \noalign{\smallskip}

ID&l    &b    &SpC$^{a}$& $v\sin i^{b}$&$v_{\rm mac}^{c}$&$T_{\rm eff}$&$\log g^{d}$&l& $\xi$        &l&$Y_{\rm He}$&l&$\log Q$&$q$&Ref.\,file&SNR\\
  &[deg]&[deg]&         &[km\,s$^{-1}$]&[km\,s$^{-1}$]   &[K]          &            & &[km\,s$^{-1}$]& &            & &        &   &          &   \\

 \noalign{\smallskip}
 \hline
 \noalign{\smallskip}		
 \endfirsthead

\caption[]{Results of the spectroscopic analysis of the stars in the sample (extract)}\\ %
 \hline\hline
 \noalign{\smallskip}		

ID&l    &b    &SpC$^{a}$& $v\sin i^{b}$&$v_{\rm mac}^{c}$&$T_{\rm eff}$&$\log g^{d}$&l& $\xi$        &l&$Y_{\rm He}$&l&$\log Q$&$q$&Ref.\,file&SNR\\
  &[deg]&[deg]&         &[km\,s$^{-1}$]&[km\,s$^{-1}$]   &[K]          &            & &[km\,s$^{-1}$]& &            & &        &   &          &   \\

 \noalign{\smallskip}
 \hline
 \noalign{\smallskip}
 \endhead

 \noalign{\smallskip}
 \hline
 \endfoot

\noalign{\smallskip}
\noalign{\smallskip}
HD~164032 & 0.8767 & -3.237 & B1/2Ib & 108 & 50 & 23300~$_{-800}^{+700}$ & 2.9~$_{-0.1}^{+0.1}$ & = & 18 & = & 0.1~$_{-0.0}^{+0.0}$ & < & -14.0~$_{-0.0}^{+0.3}$ & 2 & HD164032\_20220830\_201330\_M\_V85000\_log & 106 \\\noalign{\smallskip}
HD~164019 & 1.9099 & -2.6166 & O9.5IVp & 78 & 46 & 31000~$_{-300}^{+300}$ & 3.3~$_{-0.1}^{+0.0}$ & = & 15 & = & 0.1~$_{-0.0}^{+0.0}$ & = & -13.6~$_{-0.1}^{+0.2}$ & 1 & HD164019\_20080608\_075144\_F\_V48000 & 183 \\\noalign{\smallskip}
HD~163613 & 2.0881 & -1.9638 & B1I/II & 71 & 78 & 24400~$_{-700}^{+600}$ & 2.9~$_{-0.1}^{+0.1}$ & = & 20 & = & 0.1~$_{-0.0}^{+0.0}$ & = & -13.5~$_{-0.2}^{+0.2}$ & 1 & HD163613\_20200804\_213522\_N\_V25000 & 96 \\\noalign{\smallskip}
HD~160430 & 3.7823 & 3.5878 & B2II & 20 & 46 & 22799~$_{-500}^{+500}$ & 3.3~$_{-0.1}^{+0.1}$ & = & 13 & = & 0.1~$_{-0.0}^{+0.0}$ & < & -13.9~$_{-0.1}^{+0.1}$ & 3 & HD160430\_20200802\_220513\_N\_V25000 & 96 \\\noalign{\smallskip}
HD~164741 & 5.164 & -1.6449 & B1III$^{(2)}$ & 20 & 32 & 25099~$_{-1200}^{+1400}$ & 3.6~$_{-0.2}^{+0.2}$ & = & 8 & = & 0.2~$_{-0.0}^{+0.0}$ & < & -14.0~$_{-0.0}^{+0.4}$ & 1 & HD164741\_20180731\_223907\_M\_V85000 & 38 \\\noalign{\smallskip}
HD~173502 & 5.3641 & -12.2722 & B0.5III$^{(2)}$ & 131 & 22 & 26600~$_{-900}^{+600}$ & 3.6~$_{-0.1}^{+0.1}$ & = & 13 & = & 0.2~$_{-0.0}^{+0.0}$ & < & -14.0~$_{-0.0}^{+0.3}$ & 3 & HD173502\_20190710\_003102\_N\_V25000 & 93 \\\noalign{\smallskip}
HD~156779 & 5.4256 & 10.3249 & B2II & 97 & 39 & 18100~$_{-900}^{+600}$ & 3.0~$_{-0.1}^{+0.1}$ & = & 13 & = & 0.1~$_{-0.0}^{+0.0}$ & < & -14.0~$_{-0.0}^{+0.4}$ & 1 & HD156779\_20210621\_224538\_M\_V85000\_log & 80 \\\noalign{\smallskip}
HD~168941 & 5.821 & -6.3128 & O9.5IVp & 107 & 91 & 30200~$_{-300}^{+400}$ & 3.2~$_{-0.0}^{+0.0}$ & = & 16 & = & 0.1~$_{-0.0}^{+0.0}$ & < & -14.0~$_{-0.0}^{+0.2}$ & 1 & HD168941\_20060512\_083253\_F\_V48000 & 363 \\\noalign{\smallskip}
HD~165016 & 5.8521 & -1.5791 & B0V$^{(2)}$ & 19 & 21 & 30299~$_{-300}^{+500}$ & 3.9~$_{-0.1}^{+0.1}$ & = & 6 & = & 0.1~$_{-0.0}^{+0.0}$ & > & -13.0~$_{-0.2}^{+0.1}$ & 1 & HD165016\_20200803\_211628\_N\_V25000 & 109 \\\noalign{\smallskip}
HD~168750 & 6.1891 & -5.8526 & B1Ib & 33 & 50 & 25600~$_{-700}^{+600}$ & 3.3~$_{-0.1}^{+0.1}$ & = & 13 & = & 0.2~$_{-0.0}^{+0.1}$ & < & -13.9~$_{-0.1}^{+0.3}$ & 1 & HD168750\_20220826\_215128\_M\_V85000\_log & 84 \\\noalign{\smallskip}
HD~149757 & 6.2812 & 23.5877 & O9.2IVnn & 410 & 0 & 30700~$_{-600}^{+600}$ & 3.3~$_{-0.1}^{+0.1}$ & = & 16 & = & 0.2~$_{-0.0}^{+0.0}$ & = & -13.0~$_{-0.6}^{+0.1}$ & 1 & HD149757\_20210622\_234421\_M\_V85000\_log & 435 \\\noalign{\smallskip}
HD~164018 & 6.6528 & 0.1591 & B1/2Ib & 142 & 38 & 29400~$_{-800}^{+600}$ & 3.6~$_{-0.1}^{+0.1}$ & = & 15 & = & 0.1~$_{-0.0}^{+0.1}$ & < & -14.0~$_{-0.0}^{+0.4}$ & 1 & HD164018\_20180920\_215654\_N\_V25000 & 70 \\\noalign{\smallskip}
HD~165132 & 6.7522 & -1.2103 & O9.7V$^{(2)}$ & 127 & 20 & 32500~$_{-700}^{+1100}$ & 4.0~$_{-0.1}^{+0.2}$ & = & 12 & = & 0.1~$_{-0.0}^{+0.0}$ & < & -14.0~$_{-0.0}^{+0.4}$ & 1 & HD165132\_20190709\_222717\_N\_V67000 & 73 \\\noalign{\smallskip}
HD~164971 & 6.8868 & -0.9279 & B0Ia & 43 & 76 & 27500~$_{-700}^{+800}$ & 3.1~$_{-0.1}^{+0.2}$ & = & 16 & = & 0.1~$_{-0.0}^{+0.1}$ & < & -13.6~$_{-0.4}^{+0.2}$ & 1 & HD164971\_20110902\_222249\_M\_V85000 & 60 \\\noalign{\smallskip}
HD~163892 & 7.1516 & 0.6161 & O9.5IV(n) & 216 & 0 & 31600~$_{-500}^{+600}$ & 3.5~$_{-0.1}^{+0.1}$ & = & 12 & = & 0.1~$_{-0.0}^{+0.0}$ & < & -13.9~$_{-0.1}^{+0.2}$ & 1 & HD163892\_20210623\_001114\_M\_V85000\_log & 142 \\\noalign{\smallskip}
HD~164402 & 7.1621 & -0.0339 & B0Ib$^{(3)}$ & 54 & 86 & 28900~$_{-400}^{+300}$ & 3.2~$_{-0.1}^{+0.1}$ & = & 18 & = & 0.1~$_{-0.0}^{+0.0}$ & = & -13.6~$_{-0.1}^{+0.1}$ & 1 & HD164402\_20220829\_203450\_M\_V85000\_log & 144 \\\noalign{\smallskip}
HD~164637 & 7.3435 & -0.2284 & B0Ib/II & 35 & 72 & 29200~$_{-200}^{+200}$ & 3.3~$_{-0.0}^{+0.0}$ & = & 15 & = & 0.1~$_{-0.0}^{+0.0}$ & < & -13.9~$_{-0.1}^{+0.2}$ & 1 & HD164637\_20130821\_042119\_F\_V48000 & 327 \\\noalign{\smallskip}
HD~164359 & 7.6967 & 0.338 & B1II & 77 & 35 & 30099~$_{-400}^{+500}$ & 3.9~$_{-0.1}^{+0.1}$ & = & 9 & = & 0.1~$_{-0.0}^{+0.0}$ & < & -14.0~$_{-0.0}^{+0.3}$ & 2 & HD164359\_20220830\_210910\_M\_V85000\_log & 131 \\\noalign{\smallskip}
HD~158661 & 8.2908 & 9.0476 & B0II & 60 & 96 & 26900~$_{-500}^{+400}$ & 3.0~$_{-0.1}^{+0.1}$ & = & 22 & = & 0.1~$_{-0.0}^{+0.0}$ & = & -13.3~$_{-0.1}^{+0.2}$ & 1 & HD158661\_20210623\_010823\_M\_V85000\_log & 151 \\\noalign{\smallskip}
HD~165812 & 8.476 & -1.1091 & B1/2II & 42 & 34 & 25099~$_{-700}^{+700}$ & 3.7~$_{-0.1}^{+0.1}$ & = & 8 & = & 0.1~$_{-0.0}^{+0.0}$ & < & -13.9~$_{-0.1}^{+0.2}$ & 1 & HD165812\_20200803\_215831\_N\_V25000 & 113 \\\noalign{\smallskip}
HD~166852 & 8.5074 & -2.3235 & B0Ia/ab & 45 & 130 & 32000~$_{-1200}^{+1400}$ & 3.4~$_{-0.2}^{+0.2}$ & = & 16 & = & 0.1~$_{-0.0}^{+0.1}$ & < & -14.0~$_{-0.0}^{+0.5}$ & 1 & HD166852\_20190709\_220222\_N\_V67000 & 46 \\\noalign{\smallskip}
HD~159864 & 8.5231 & 7.3825 & B1Ib & 88 & 98 & 27300~$_{-400}^{+600}$ & 3.1~$_{-0.1}^{+0.1}$ & = & 18 & = & 0.1~$_{-0.0}^{+0.0}$ & < & -13.7~$_{-0.3}^{+0.2}$ & 1 & HD159864\_20200502\_050638\_N\_V25000 & 110 \\\noalign{\smallskip}
HD~165516 & 8.9268 & -0.4446 & B1/2Ib & 47 & 67 & 25900~$_{-400}^{+300}$ & 3.1~$_{-0.1}^{+0.1}$ & = & 17 & = & 0.1~$_{-0.0}^{+0.0}$ & = & -13.7~$_{-0.2}^{+0.2}$ & 1 & HD165516\_20200803\_212544\_N\_V46000 & 144 \\\noalign{\smallskip}
HD~165892 & 9.1729 & -0.8113 & B2II & 74 & 42 & 21600~$_{-900}^{+1000}$ & 3.4~$_{-0.2}^{+0.1}$ & = & 11 & = & 0.1~$_{-0.0}^{+0.0}$ & < & -14.0~$_{-0.0}^{+0.3}$ & 1 & HD165892\_20130412\_050607\_M\_V85000 & 62 \\\noalign{\smallskip}
HD~149363 & 9.8524 & 26.6906 & B1/2Ib & 88 & 62 & 27300~$_{-200}^{+400}$ & 3.3~$_{-0.1}^{+0.1}$ & = & 16 & = & 0.1~$_{-0.0}^{+0.0}$ & < & -14.0~$_{-0.0}^{+0.3}$ & 1 & HD149363\_20190707\_211718\_N\_V25000 & 194 \\\noalign{\smallskip}
HD~164438 & 10.3529 & 1.7886 & O9.2IV & 61 & 105 & 31300~$_{-400}^{+400}$ & 3.3~$_{-0.0}^{+0.0}$ & = & 13 & = & 0.1~$_{-0.0}^{+0.0}$ & < & -14.0~$_{-0.0}^{+0.5}$ & 1 & HD164438\_20080514\_092403\_F\_V48000 & 303 \\\noalign{\smallskip}
HD~166546 & 10.358 & -0.9242 & O9.5IV & 31 & 73 & 31000~$_{-200}^{+400}$ & 3.4~$_{-0.0}^{+0.0}$ & = & 12 & = & 0.1~$_{-0.0}^{+0.0}$ & < & -14.0~$_{-0.0}^{+0.1}$ & 1 & HD166546\_20080609\_082259\_F\_V48000 & 284 \\\noalign{\smallskip}
HD~167264 & 10.4557 & -1.7408 & O9.7Iab & 80 & 67 & 28200~$_{-300}^{+300}$ & 3.1~$_{-0.0}^{+0.0}$ & = & 22 & = & 0.1~$_{-0.0}^{+0.0}$ & = & -13.2~$_{-0.0}^{+0.0}$ & 1 & HD167264\_20100907\_202837\_N\_V46000 & 273 \\\noalign{\smallskip}
... & ... & ... & ... & ... & ... & ... & ... & ... & ... & ... & ... & ... & ... & ... & ... & ... \\
\end{longtable}

\tablefoot{
\tablefoottext{a}{We adopted the recommended classification quoted in the \textit{Simbad} astronomical database \citep{wenger00} except for the cases indicated with an upper index and the following: ($1$) - \citet{deburgos20}; ($2$) - \citet{deburgos23}; ($3$) - \citet{negueruela-subm}. For HD~168021, HD~168183, and HD187879, the \textit{Simbad} classification included a second component, but we have only included the first. We note that in the case of the O-type stars, most of the classifications are adopted from the GOSSS series of papers \citep{sota11,sota14,maiz-apellaniz16}}.
\tablefoottext{b}{We adopt a global uncertainty for \vsini of 15\kms based on the average value.}
\tablefoottext{c}{We adopt a global uncertainty for \vmac of 20\kms based on the average value. Based on the results by \citet{simon-diaz14a}, we assigned \vmac=\,0\kms to those cases in which \vsini>\,180\kms.}
\tablefoottext{d}{The values of \logg have not been corrected for centrifugal acceleration.}
The full table is available at the CDS.
}

\end{landscape}
}

\end{appendix}

\end{document}